\documentclass[11pt]{article}
\usepackage{geometry}                
\usepackage{graphicx}
\usepackage{amssymb}
\usepackage{epstopdf}
\title{QCD and QED dynamics of the EMC effect}
\author{ Leonid Frankfurt, \\ School of Physics and Astronomy, Tel Aviv University, 69978 Tel Aviv, Israel \\
\\
Mark Strikman \\
Penn State University, University Park, PA, 16802, U.S.A.}

\date{}
\begin{document}
\maketitle

\begin{abstract}
Applying exact QCD sum rules for the baryon charge and energy-momentum we demonstrate  that if  nucleons  are the 
only degrees of freedom of nuclear wave function ,   the structure function of a nucleus would be the   additive sum of the nucleon distributions at the same Bjorken  $x=AQ^2/2(p_A\cdot q)\le 0.5$ up to very small  Fermi motion corrections if  $1/2m_{N} x$ is significantly less than the nucleus radius. Thus observations at CERN, SLAC, and TJNAF of the deviation of 
the ratio $R_A(x,Q^2)=(2/A)F_{2A}(x,Q^2)/F_{2D}(x,Q^2)$ from one reveal  the presence of non-nucleonic  degrees of 
freedom in nuclei.  Employing the parton model (the QCD evolution equation) or exact QCD sum rules  shows 
that the ratio  $R_A(x_{p},Q^2)$  used in experimental studies,   where  $x_p=Q^2/2q_0m_p$ deviates  from one even if a 
nucleus consists of nucleons with small momenta only.   Use of the Bjorken $x$ in the theoretical analysis of experimental data  
leads in  the case of the light nuclei to additional decrease of $R_A(x,Q^2)$  as compared to the plots presented
in the experimental papers.  Coherent contribution of Fermi, Weizsacker, Williams equivalent photons into photon  
component of parton wave function of a nucleus unambiguously follows from Lorentz  transformation of the rest frame 
nucleus Coulomb field. Account of  light cone fraction of nucleus momentum  carried by equivalent photons  almost compensates the difference between 
data analysis in terms of Bjorken $x$ and $x_p$ for heavy nuclei. $Q^2$ dependence of the hadronic  EMC effect  
emphases  difference of the interplay of the leading twist and higher twist effects  for $Q^2$ probed at  TJNAF and SLAC energies and those probed at CERN. Direct  observations of large and predominantly nucleonic short-range correlations 
(SRC) in nuclei pose a serious challenge for most of the proposed models of the EMC effect for $x\ge 0.6$.  The data are consistent with a scenario  in which  the hadronic  EMC effect reflects 
fluctuations of inter nucleon interaction due to fluctuations of color distribution in  the interacting nucleons. 
The dynamic realization of this scenario is presented in which quantum  fluctuations of the nucleon wave  function with 
$x\ge 0.5$ parton have a weaker interaction with nearby nucleons, leading to suppression of such configurations in 
bound nucleons and to the significant suppression of  nucleon Fermi motion effects  at $x\ge 0.55$ giving a right magnitude of the EMC effect. The directions for the future studies and challenging questions are outlined.\end{abstract}

\section{Introduction}

Nearly thirty years ago the  first observation of a difference between the nucleon and nucleus parton distributions was 
reported by the European Muon Collaboration (EMC) \cite{EMC}.    Hence such a  difference is referred to as the EMC 
effect. The EMC measurement was  followed  by a series of  experiments in the eighties which provided first information 
on the A-dependence of the EMC effect \cite{SLAC} and the   A-dependence of antiquark distributions \cite{DY}. For a 
review of the data collected during the first decade of 
studies of the nuclear parton distributions see \cite{dataa}.   Interest in the EMC effect was  revitalized by the recent high precision measurements at TJNAF for a range of the lightest nuclei \cite{Seely:2009gt}. To extract the EMC effect from experimental data one should account   properly  for  the QCD dynamics of hard processes 
and the QED physics of equivalent photons accounting for electric charge of nucleus.

The parallel development  is the BNL and TJNAF  experimental studies which allowed one to observe directly  the short-range 
correlations (SRC) in nuclei and explore their structure, for a review see \cite{Frankfurt:2008zv,Arrington:2011xs}. When combined with other experimental measurement of hard nuclear phenomena these observations put very strong constraints 
on the interpretation of the origin of the EMC effect which are not satisfied by many of the proposed 
explanations of the effect.

Another challenge  for  the models of the EMC effect based on the ideas taken from low energy nuclear physics are the
data on antiquark distribution within a nucleon, nucleus.

In the review we restrict our discussion to  the region   $x\ge 0.2$
where nuclear shadowing (anti shadowing)  phenomena are 
not important. For the recent review of small $x$ leading twist physics and extensive list of references see 
\cite{Frankfurt:2011cs}.

Within the leading twist approximation  of QCD  nuclear structure  functions depend on the  Bjorken variable 
$x=AQ^2/2(q\cdot p_A)$ and $Q^2$ as the direct consequence of the 
dominance of hard interaction with a single parton in the leading twist hard processes. 
Description of structure functions in terms of the Bjorken variable $x$ and $Q^2$ accounts for the 
renormalizability of QCD and QED.  Account of Bjorken $x$ i.e. the QCD dynamics removes artificial
effect introduced by the EM collaboration and followed by all other experimental groups  who ignored parton structure 
of wave function of nuclear target by using non-partonic variable $x_p=Q^2/2q_0m_p$ to compare  nucleus and deuteron structure functions,  see Figs.~1 and 2. Note that account of this effect leads to certain decrease of structure function of a nucleus  i.e. sign of this effect is opposite to the one claimed in \cite{Frankfurt:2010cb}.

In the case of medium and heavy nuclei  an additional effect becomes important:   a nucleus at rest has an electric charge $Z$ 
and related  Coulomb field which is the zero component of electromagnetic field. Under the  Lorentz transformation to the 
frame where the nucleus  has a large momentum, the  nucleus Coulomb field  is transformed into the field of equivalent photons. This phenomenon is well known as  the Fermi-Weizsacker-Williams (FWW) approximation \cite{WW}.  Equivalent photons carry a noticeable and calculable fraction of the nucleus momentum if the parameter characterizing QED 
phenomena $Z\alpha_{em}$  is not  too small.   We evaluate this fraction which is  dominated by the coherent contribution 
into photon distribution in a heavy nuclear target \cite{Frankfurt:2010cb}.

Incoherent contribution into photon component of nucleus wave function is not negligible but it produces practically the 
same effects in the  nucleus and nucleon  distributions.  So corrections to the additivity of nucleon structure functions 
due to incoherent photon component are tiny:  $\approx  (Z/A)\alpha_{em}$  multiplied by the probability of admixture of
non-nucleon degrees of freedom in the nucleus wave function. The coherent photon distribution in protons and neutrons   
was   evaluated previously in  \cite{Kniel,Gluck:1994vy}, while in \cite{Gluck:2002fi} the $ Q^2$ evolution of incoherent
photon distribution within a nucleon resulting from the $Q^2$ evolution of quark, antiquark nucleon parton distribution 
functions (pdfs)  has been calculated \footnote{Authors did not explicitly considered the total momentum fraction, though 
they calculated the  photon parton distributions for all $x$.}. However the implications of photon parton component for the momentum sum rule -- one of the important phenomena  in the case of medium and heavy nucleus -- was 
considered only in \cite{Frankfurt:2010cb}.

Discussed above experimental and theoretical observations gives us a clue to  the physics relevant for superdense 
nuclear matter - 3--5 nuclear densities (inner core of neutron stars, etc.)   Really the BNL and TJNAF data show that 
short range nucleon correlations (SRC) in nuclei are dominated by the nucleon degrees of freedom; lack of nuclear 
effects in the antiquark distribution of nuclei shows that  meson fields of bound and free nucleons  are close at least 
up to the densities characteristic for SRC i.e. $\sim $ 3--5 average  nuclear densities.  Electromagnetic radius of bound and 
free nucleons are close.  All these observations support validity of the equation of state suggested by the  text book 
nuclear theory and therefore consistent with the recent observation of neutron stars with the mass around two  solar 
masses. More generally these facts indicate that transition from low scale dynamics  successfully described by effective 
Chiral  QCD Lagrangian ($\Lambda_{\chi} \sim {\mbox 1 GeV}$) to dynamics at higher resolution scale should be pretty intricate to be consistent  with data on antiquark distribution within nucleons and nuclei.

The review is organized as  following.   In  section 2 we explain why and how an account of the QCD dynamics 
allows us to fix dynamic variables eliminating  kinematical effects introduced in the experimental studies 
which lead to the artificial  breakdown  of the nucleon additivity for the nuclear structure functions in the approximation 
when the nucleus is built of nucleons only. 
 
In  section 3 we build and analyze standard nonrelativistic nuclear physics approach  where 
internal nucleon motion and the c.m. motion of the nucleus are independent.  Within this approach a 
nucleus in the infinite momentum frame  (or light-cone)  consists of nucleons only. We use  the exact  QCD  baryon 
number  and momentum sum rules to derive the formula for the nuclear pdfs which takes into 
account nucleon  Fermi motion  and show that corrections to  the ratio of nuclear structure functions due to this effect 
are small for $x \le 0.55 (0.7) $ for $F_{2N}(x,Q^2) \propto (1-x)^3 (\propto  (1-x)^2)$.  Hence it will be  convenient in our theoretical discussion to consider separately two kinematical regions:   $x\le 0.5$ where the answer is expressed through 
the average deformation of the structure functions of the bound nucleon and $x\ge 0.6$ where interplay of quantum 
fluctuations of nucleon properties  with the nucleon Fermi motion effects and SRC  sets in. 

In  section 4 we evaluate the contribution of the simplest non-nucleonic component of nuclear structure functions - the 
photon structure function  of a nucleus, $P_{A}(x,Q^2)$,  calculate the fraction of the nucleus  momentum carried 
by equivalent photons and extend the standard nuclear model to include this effect.  This is feasible since
coherent effects which give the dominant contribution in this calculation are model independent.

In section 5 we consider interplay of two  model independent effects discussed in sections 3 and 4. 
EM collaboration introduced artificial correction to the ratio of nuclear and nucleon structure functions  by comparing 
structure functions  at $x$ variable which differ  from conventional Bjorken variable $x$ required by QCD dynamics.  
Removing this  correction  leads to the increase of the EMC effect by about 20\% for $^4$He   and a factor of two 
smaller for heavy nuclei (the absolute correction is about the same for $^4$He while the EMC effect grows with A).
At the same time the Coulomb effect discussed in section 4 generates a EMC like effect of comparable strength for 
$A\sim 200$  but opposite sign.  As a result the hadronic component of the EMC effect is well described by the EMC 
ratio for heavy nuclei but underestimated for light and medium nuclei. The remaining effect represents the genuine 
"hadronic" EMC ratio which should be compared with the expectation of the Fermi motion contribution. One observes 
that deviation from the Fermi motion does not exceed 5\%  for $x \sim 0.5$ and much smaller for smaller x, indicating 
that    wave functions of bound and free nucleon are very close for most of the quark-gluon configurations  even at the 
central nuclear densities.

In  section \ref{section6} we summarize  the recent  direct observations of the short-range correlations in nuclei which 
indicate that probability of such correlations is large (on the scale of 20\%) and that nucleons in the SRCs are rather weakly deformed.

In  section \ref{section7}  we discuss interrelation between SRC and EMC effect and demonstrate that the A-dependence of the hadronic component of the EMC effect is consistent with the A-dependence of the SRCs.  

In section \ref{section8}  we summarize the most important constraints on the models of the EMC effect coming from the experimental studies of hard phenomena with nuclei and explain  that wide classes of the current models are not consistent 
at least with one of these constraints. 

In section \ref{section9} we consider plausible interrelation between  well established properties of hard processes off
a free nucleon and   dynamics of the hadronic EMC effect. We explain that the   observed rapid decrease of antiquark 
distribution  within a nucleon at $x\ge 0.4$  indicates the suppression of meson field of a bound nucleon participating in 
hard processes.   We argue that this suppression leads to a significant  suppression of the contribution of SRC in the 
hard processes which select certain quark-gluon configuration in an interactive nucleon.Therefore Fermi motion effects  
and hence nuclear pdfs at $x\ge 0.6$ are suppressed as compared to the standard nuclear theory giving a right magnitude 
of the hadronic EMC effect at large $x$.

 In section 10 we discuss briefly implications of discussed in the review effects on global fits of pdfs.

In section 11 we  present the conclusions and implications of the EMC effect related physics  for different phenomena.

\section{Account of the  QCD dynamics  fixes  the kinematic 
breakdown of the nucleon additivity for nuclear pdfs}
\label{section2}

It was proved long ago that in the limit $Q^2\to \infty$, where $Q^2$  is the square of four-momentum transferred 
by electrons to the target, $T$,
but fixed  Bjorken $x_T$, the  fraction of nucleus momentum carried by interacting parton:
\begin{equation}
x_T=Q^2/2(q\cdot p_T),
\label{Bj0}
\end{equation}
the structure functions of any target T are given by the convolution of cross section for hard probe scattering off individual 
patrons with parton distribution within this target:
\begin{equation}
\sum_n \int \psi_n^2(x_i,Q^{2}_0) \prod (dx_i)\delta (1-\sum x_i) \delta (x_T-x_i).
\end{equation}
The Bjorken scaling for the  structure functions and $Q^2$ evolution follow  from the asymptotic freedom in QCD and is 
implemented  within  the parton model accounting for the $Q^2$ evolution, the light-cone quantization of QCD and the 
Wilson operator  product expansion.   Normalization of the light-cone wave function (WF) of a target is derived based 
on the evaluation of the matrix elements of the exactly conserved currents between the  WFs of the target in initial and 
final states  at the zero momentum transfer (electromagnetic current, charmed, bottom currents).  The parton model  generalized to include the $Q^2$  evolution is the basis  for QCD physics of hard processes, for the searches  of new 
particles, etc. 

Thus the appropriate variables to describe the process $e+A\to e'+X$ in terms of nuclear pdfs 
are the Bjorken variable $x_A$ ($\equiv x_T$)  and $Q^2$.   To simplify formulae it  is
convenient to rescale Bjorken $x_A$ by the factor $A$:
\begin{equation}
x/A=Q^2/(2q\cdot p_A).
\label{Bj1}
\end{equation}
So $x/A$ is the fraction of the total nucleus momentum carried by the interacting parton and $0< x/A <1$. 
To investigate nuclear effects which can be interpreted within QCD 
the EM Collaboration  introduced the ratio:
\begin{equation}
R_A=(2/A)F_{2 A}(x,Q^2)/F_{2D}(x,Q^2).
\label{additive1}
\end{equation}

It is important to realize that the EMC ratio as defined in Eq.\ref{additive1}   is presented in the experimental papers 
as a function of the  variable $x_p=Q^2/2q_0m_p$ \cite{EMC} (also it was normalized  to the  cross section of electron -- deuteron scattering rather than to the sum of the electron -- proton and electron--neutron cross sections).  Such variable 
is convenient for  experimental studies as one can compare the cross sections of the lepton - nucleus scattering for 
the same kinematics of the  incident and final lepton. Variable $x_p$ differs from the parton model variable $x$ and 
therefore within the parametrizations of structure functions in terms of $x_p$  Bjorken  scaling and  well established 
universality of hard processes obtain unconventional form which is in variance with the physical intuition.
In particular,   the ratio $R_A(x_p,Q^2)$ defined similar to Eq. \ref{additive} is different from one  in the kinematics where 
$\gamma^*$ scatters off bound nucleon carrying $x_A=1/A$ fraction of nucleus momentum, i.e. off a bound nucleon at rest in the nucleus target rest frame.   Thus the baryon charge and momentum sum rule are violated 
within a model where nucleus consists of nucleons only etc.       Moreover nuclear pdfs  presented as a function of $x_p$ 
cannot be used  directly without the additional correcting factor to predict hard phenomena in $pA$ and $AA$ collisions 
where as the consequence of the  QCD factorization theorem the hard cross sections are controlled by the parton 
distributions over the Bjorken x fraction of the colliding nucleus energy carried by the interacting parton.

We will demonstrate in the next section that the use of the Bjorken $x$ instead of $x_p$  leads to a $~ 20 \%$  enhancement 
of the EMC effect for the lightest nuclei ($A\le 12$) for $x\sim 0.5$ and to somewhat smaller correction for heavy nuclei. 
We correct here mistake made in \cite{Frankfurt:2010cb} in the process of restoring Bjorken scaling and QCD evolution 
violated in experimental studies, see also \cite{Heli} and Fig.\ref{xscale} below. 
For medium and heavy nuclei a comparable contribution to the EMC ratio but of opposite sign
originates from  another model independent effect - presence of the equivalent  photon 
field of nuclei (section \ref{section4}).

\section{Standard model for the structure functions of nuclei}
\label{section3}

In order to discuss to what extent  the EMC effect signals presence of new physics in the nuclear structure we need 
to establish expectations of the textbook nonrelativistic  nuclear model  in which  nuclei are build  only of nucleons 
which have the same internal structure as free nucleons.  We will refer to this approximation as the standard nuclear 
model.   (Actually the standard model  should be extended  to include also
equivalent photons as non-nucleonic degrees of freedom. This will be done in the next section.)

At the first step  of our analysis the Fermi motion of nucleons  can be neglected. Let us consider a nucleus moving 
with a large momentum $P$ and a  parton which  carries a fraction $x_A$   of the nucleus momentum. Each nucleon 
carries fraction $P/A$ of the nucleus momentum  since the Fermi motion  of the  nucleons
is neglected. Hence this parton carries a fraction $Ax_A$ of the nucleon momentum. As a result we find
\begin{equation}
f^{j}_A(x_A,Q^2) = Z f^{j}_p(Ax_A,Q^2) + N f^{j}_n(Ax_A,Q^2),
\label{additive}
\end{equation}
where $x_A=Q^2/2(q\cdot P)$.  In the case of deep inelastic scattering necessity to use $x_A$ follows from the 
fundamental property of parton model that $\gamma^*$ interacts with an individual parton.  Note that the  exact 
QCD sum rules for the baryon charge and momentum conservation were derived using Eq.\ref{additive} or by 
direct calculation of matrix elements of conserved currents: the baryon charge and the energy-momentum tensor. 
Since in this approximation $f^{j}_A$ is equal zero for $x_A\ge 1/A$ it is convenient to rescale 
\begin{equation}
x_A \to Ax_A=x, 
\end{equation}
so that $x$ is now changing between 0 and A and consider the ratio 
\begin{equation}
R^{j}_A(x,Q^2)= {f^{j}_A(x,Q^2) \over  Z f^{j}_p(x,Q^2) + N f^{j}_n(x,Q^2)}.
\label{rgdef}
\end{equation}
If there are no nuclear effects,
\begin{equation}
R^{j}_A(x,Q^2) =1.
\end{equation}

In the nonrelativistic approximation to  the  nucleon Fermi motion, the nucleus c.m. motion is separated from the inner motion  
of nucleons.  So  the light- cone WF of a nucleus, $\psi_A=\exp i(\vec{p}_A\cdot {\vec x})\psi_{int}$, can be conveniently  calculated  in terms of equal time WF of the nonrelativistic nuclear theory - $\psi_{int}$.   Here $p_A$ is nucleus momentum  and $\vec{x}$ the coordinate of the nucleus center of mass. Thus the  Lorentz boost in 
 this approximation is trivial -- it is reduced to the  transformation of the plane wave. 
 This property allows us to use the experience of  the low energy nuclear physics to evaluate some hard QCD phenomena.

There exists a variety of the exact sum rules for valence quark distribution which follow from the Ward identities and 
existence of exactly conserved currents: baryon charge, isotopic charge etc.  The typical valence quark sum rule for the
baryon charge and the momentum sum rule are:
\begin{equation}
\int^A_0 dx V_A(x,Q^2)=B,
\label{vsumrule}
\end{equation}

\begin{equation}
\int^A_0  dx[xV_A(x,Q^2)+xS_A(x,Q^2)+xG_A(x,Q^2)]=A.
\label{msumrule}
\end{equation}
Here $xV_A(x,Q^2)$ is the density of valence quark distribution,  $x_{A}S_A(x_A,Q^2)$ is the density of nucleus sea quark distribution,  $xG_A(x,Q^2)$ is the density of nucleus gluon distribution.  $B$  is nucleus baryon charge. An additional factor of 
$A$ in  Eq. \ref{msumrule} reflects our choice of scale for $x$ -- the fast momentum is measured in units of $p_A/A$ that is 
average nucleon momentum (in the model where nucleus is build only of nucleons in the fast frame).

In this section we consider nucleus as a system of nucleons and use  the above sum rules to derive approximate formulae for 
the nuclear pdfs accounting for relativistic corrections due to nucleon Fermi motion. In the impulse approximation   nuclear pdfs  
are described by the convolution formulae :
\begin{equation}
 f^{j}_A(x,Q^2)= \int_x^A {d\alpha\over \alpha} \left[ f^{j}_p(x/\alpha,Q^2)\rho^p_A(\alpha) +f^{j}_n(x/\alpha,Q^2)\rho^n_A(\alpha)\right]. 
 \label{nucpdf}
\end{equation}
Here $\rho^{p,n}_A(\alpha)$ is the proton (neutron) light-cone densities of the  nucleus,  $\alpha/A$ is the fraction of nucleus momentum carried by interacting nucleon, $Z$ and $N$ are  the numbers of protons and neutrons. The evaluation of the matrix element of the conserved currents (electromagnetic, isotopic, baryon charge) 
 -- $\left<A\right|J_{\mu}(t)\left|A\right>$
at the zero momentum transfer, $t$ between nucleus states  gives \cite{Frankfurt:1981mk}:
\begin{equation}
\int^A_0  \rho_{A}^{p,n}(\alpha) {d\alpha\over \alpha} =Z(N).
\label{CHARGE}
\end{equation}
Evaluation of  the matrix element of the   energy-momentum tensor gives \cite{Frankfurt:1981mk}:
\begin{equation}
\int^A_0  {1\over A} [\rho_{A}^{p}(\alpha)+\rho_{A}^n(\alpha)] d\alpha=1.
\label{TEM}
\end{equation}
Since the nucleus is a nonrelativistic system, the light-cone density is concentrated near $\alpha = 1$. Hence it 
 is legitimate to evaluate the contribution of the Fermi motion for 
$x < 0.7$ by decomposing integrand of Eq.\ref{nucpdf} in the Taylor series in  powers of $\alpha-1$ and account for the first three terms in the expansion. Thus  the contribution of the Fermi motion effect to the ratio defined in Eq. \ref{rgdef}  is 
\cite{Frankfurt:1981mk,Frankfurt:1988nt}:
\begin{eqnarray}
R_A(x,Q^2)={1\over A} \int_0^A \rho_{A}^{N}(\alpha) {d\alpha\over \alpha} 
+{xf_j{^N}^{\prime}(x,Q^2)\over f_j^N(x,Q^2)} \left[1-\int_0^A \rho_{A}^{N}(\alpha) d\alpha/A\right]\nonumber \\
 +  {xf_j^{N\,\prime}(x,Q^2)+{x^2\over 2} f_j^{N \,\prime\prime}(x,Q^2)\over f_j^N(x,Q^2)} {1\over A}\int_0^A \rho_{A}^{N}(\alpha)(1-\alpha)^2 {d\alpha\over \alpha}. 
 \label{series}
\end{eqnarray}
Here $f_{j}^N=(f_{j}^p+ f_{j}^n)/2$  and we assumed equal number of protons and neutrons.  In the first term in the above equations the nucleon structure functions are cancelled in the ratio since the variable $x$ is independent of the  atomic number the resulting factor is equal to one due to the normalization condition Eq.\ref{CHARGE}.

 It is worth emphasizing that the discussed decomposition becomes inapplicable at large $x$ since it diverges for $x\sim 1$. In this region $R_A(x,Q^2)/R_{A'}(x,Q^2) \sim \langle T_A\rangle/\langle T_{A\prime}\rangle$ which is qualitatively different from Eq.\ref{series}. 
 Here $\langle T_A\rangle$ is the average kinetic energy.

It is convenient to introduce 
\begin{equation}
\eta_A=1-\int_0^A \rho_{A}^{N}(\alpha) d\alpha/A\equiv \left <1-\alpha\right>,
\end{equation}
the fraction of the  light-cone momentum which is  carried by  nuclear constituents other than nucleons. It is equal to 
zero if nucleus consists of nucleons only
, but we keep this term in anticipation of the discussion of the 
effect 
of equivalent photons
below.  In the nonrelativistic  limit  $k^2/m_N^2, \epsilon_A/m_N \ll 1$  ($\epsilon_A$ is the nuclear binding 
energy per nucleon) the nuclear factor in the third term  of Eq.\ref{series}
$ \langle(\alpha-1)^2\rangle =A^{-1}\int_0^A \rho_{A}^{N}(\alpha)(1-\alpha)^2 {d\alpha\over \alpha}$ can be calculated using  
nonrelativistic approximation for $\alpha \approx 1+k_{3}/m_{N}$ where $k$ is nucleon three momentum within the nucleus target. (The formula $\alpha \approx 1+k_{3}/m_{N}+ O(k^2/m^2_N)$    
This leads to 
\begin{equation}
\langle(\alpha-1)^2\rangle= {k^2\over 3m_N^2} + O(k^4/m_N^4, \epsilon_A^2/m_N^2)= {2T_A\over 3 m_N}.
\end{equation}
  This allows us to rewrite Eq. \ref{series} as
\begin{eqnarray}
R_A(x,Q^2)= 1 
+{xf_j{^N}^{\prime}(x,Q^2)\over f_j^N(x,Q^2)} \eta_A +  {xf_j^{N\,\prime}(x,Q^2)+{x^2\over 2} f_j^{N \,\prime\prime}(x,Q^2)\over f_j^N(x,Q^2)} {2T_A\over 3m_N}.
\label{sercon}
\end{eqnarray}

In the case of $f_{N}^j(x,Q^2) \propto (1-x)^n$, Eq. \ref{sercon} can be written as
\begin{equation}
R_A(x,Q^2) =1 + {n x  (x(n+1) -2) \over (1-x)^2}\cdot 
{T_A\over 3m_N} -\eta_A nx/(1-x).
\label{Fermimotion}
\end{equation} 
We will present numerical results in section \ref{section5}.  Here we just note that it follows from Eq.\ref{Fermimotion} 
that  for $x\le 0.5$  ($x\le 0.7$)  for n=3(n=2)
effects of Fermi motion are small and under control  as the consequence of the account for the exact conservation 
laws. The Fermi motion gives zero contribution to $R_A$ at $x=2/(n+1)$,  a small negative contribution at $x< 2/(n+1)$ 
and leads to a rapid growth of $R_A$ at $x> 2/(n+1)$.
In this discussion we ignored $Q^2$ dependence of   structure functions. We will show in the end of this section that 
higher twist (HT) effects play significant role at $\ge 0.5$ and $Q^2 \sim \mbox{few GeV}^2$.

\begin{figure}[t]  
   \centering
   \includegraphics[width=0.5\textwidth]{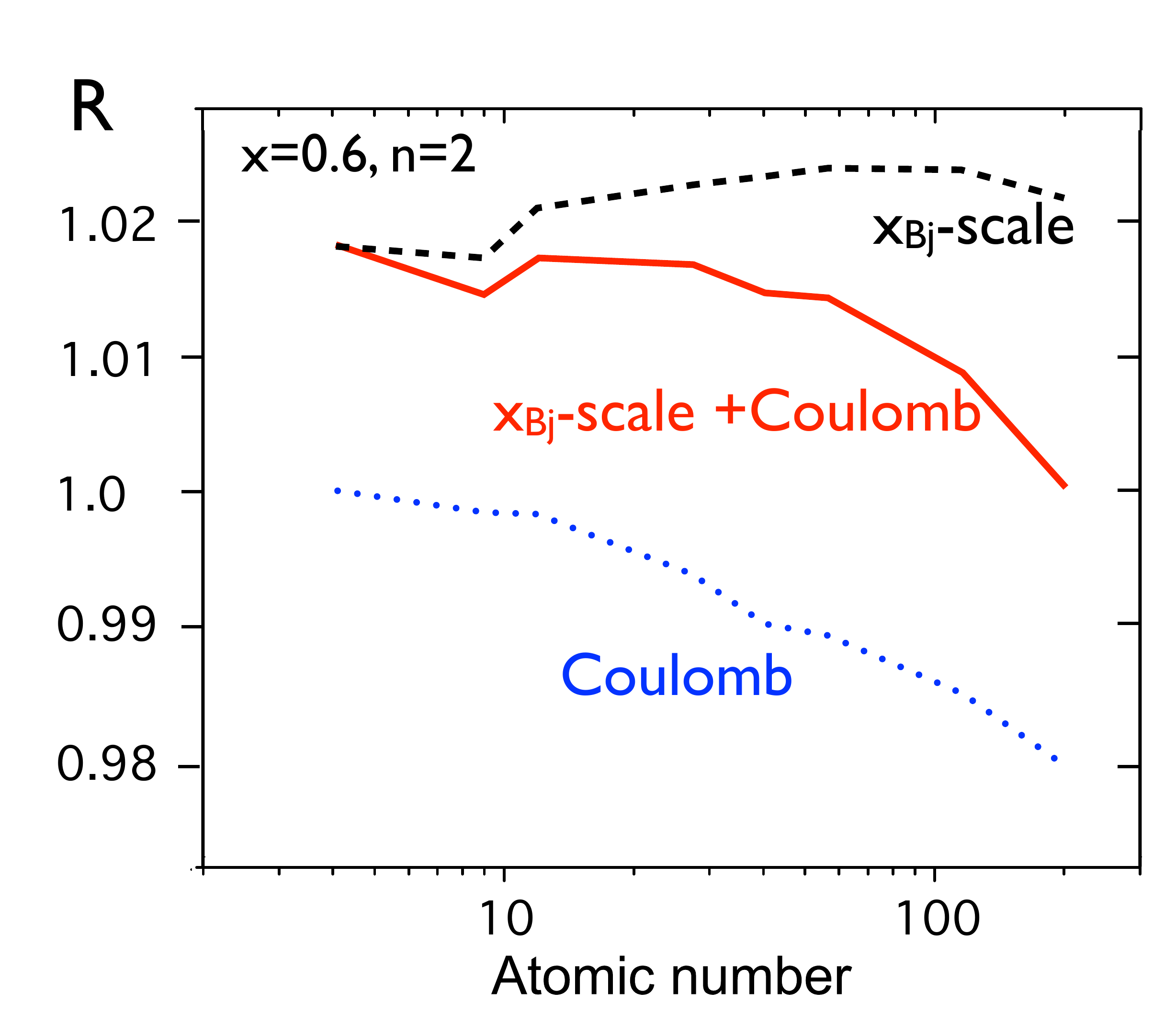} 
   \caption{Change of R due to account for correct x-scale( dashed line), 
   contribution of equivalent photons
    (dotted line) and combined effect (solid line) as a function of atomic number for $x=0.6$ and $F_{2N}(x) \propto (1-x)^2$.}
   \label{xscale}
 \end{figure}

At  the same time the experimental data on nuclear pdfs are usually presented in terms of the variable 
\begin{equation}
x_{p}= Q^2/(2m_pq_0),
\label{EMC}
\end{equation}
which is different from the fraction of target momentum carried by interacting parton and therefore requires rewriting formulae 
for parton model and $Q^2$ evolution. This variable depends on the atomic number of a target  even if the nucleon Fermi motion effects are  neglected.
\begin{equation}
x_{p}/x=m_A/Am_p= (1- (\epsilon_A - (m_n-m_p) N/A)/m_p)\equiv 1+ r_x.
\label{rx}
\end{equation}
Thus  the structure functions of interacting nucleons in the nominator and denominator depend on different arguments  \footnote{ In ref. \cite{Frankfurt:2010cb} and in the original version of this paper a mistake in the sign of $r_x$ was  made which resulted in  a wrong sign of the discussed effect.}
Thus the use of variable $x_{p}$  instead of $x$ (and the expectation  of absence of the nuclear effects in such an approximation)  violates important condition of standard nuclear model that nuclear pdfs are the sum of nucleon pdfs if the Fermi motion effects are neglected  and  introduces artificial dependence of $R_A$ on atomic number.  In particular 
$R_A(x)=1$ corresponds to
\begin{equation}
R(x_{p})= f^j_A(x(1+r_x))/f^j_N(x)\approx 1  - r_xn{x\over 1-x},
\label{rs}
\end{equation}
where at the last step we took $f^j_N(x)\propto (1-x)^n$
\footnote{Note that to simplify the expressions we took here $m_n=m_p$ so  the dominator $x$ consider with $x_p$. In 
the final expressions the A/D ratio we will   take into account the difference between  $x_p$ and $x_D$.}. The value 
of the correction to the EMC effect due to this effect is presented in Fig.\ref{xscale} together with the effect
of equivalent photons  to be discussed in the next section and combined effect which reflects the change of the 
hadronic component of the EMC effect. 
  
The nucleon Fermi motion correction can be easily included as well.   We will demonstrate  in section \ref{section5} that 
Eq.\ref{rs} leads to decrease of $R_A$, enhancing the EMC effect  for $A\ge 4$ by practically the same amount due to 
a weak A-dependence of $x_p$  for $A\ge 4$.   We will explain in the next section that for heavy nuclei   presence 
of the Coulomb field for a nucleus at rest  explains a certain  fraction of the EMC effect which compensates the effect of 
change of $x_p$. So the hadronic contribution to the EMC ratio
for heavy nuclei is close to the EMC ratio reported experimentally.   Overall account 
of the two effects leads to reduction of the A-dependence of the hadronic contribution to the EMC ratio 
for A between 4 and 200, cf. Fig. \ref{x05}.

{\it Comment.} \, A popular  expectation in the low energy nuclear physics is that one can account for the effects of relativistic 
nucleon Fermi motion  assuming  that the   vertex functions in the Feynman diagrams with a virtual nucleon coincide
with Schrodinger WFs of a nucleus. This model has been applied in several papers to explain the EMC effect
formulated in terms of non-parton model variable $x_p$ and without subtraction of the contribution of equivalent photons.
In the first papers \cite{violbsr,violbsr1}  the  baryon sum rule, i.e., the  probability conservation was violated. Later the baryon 
charge sum rule has been taken into account following the prescription of \cite{Frankfurt:1985ui}.  This 
description does not allow one  to account consistently  for the  probability conservation i.e. to satisfy simultaneously  
both  baryon and momentum sum rules for the hadronic part of the EMC effect.   
Thus within this approach light-cone nucleus wave function contains non-nucleonic degrees of freedom. 
They are hidden in the internucleon interaction and identified in the text books of nuclear physics with meson exchanges.  
Such hypothesis has problem to explain the observation of no nuclear effects in the antiquark distribution in nuclei.

The baryon sum rule in this model is derived by calculating matrix element of baryon charge exactly. 
The calculation leads to the normalization of the nucleus wave functions in terms of the value of nucleus baryon charge. 
This unambiguous normalization differs from  the nonrelativistic normalization of the nucleus WF. 
\footnote{For the  detailed discussion of the early theoretical studies of these issues and references see 
\cite{Frankfurt:1988nt}.}   (Note that normalization of wave function  natural for  the nonrelativistic physics implies
correction  to the total cross section of  DIS off deuteron of the same magnitude as the Glauber shadowing correction 
which violates probability conservation -- the so called West correction\cite{West:1972qj}).  In this approach there is no symmetry between distributions over fractions of nucleus momentum carried by interacting nucleon and nucleon 
spectators. As a result  in this model significant light-cone fraction of the nucleus momentum is carried by the 
non-nucleonic degrees of freedom \cite{Frankfurt:1988nt}:
\begin{equation}
\eta_A= \epsilon_A/m_N+ T_A/3m_N,
\label{etaa}
\end{equation}
Here $\epsilon$ is the nucleus energy binding per nucleon. $T_A$ is the average kinetic energy of a bound nucleon,
For the realistic nuclear WFs and $A\ge  40$, Eq. \ref{etaa} leads  to $\eta_A\sim 2\%$ which corresponds for  the 
Jlab, SLAC kinematics where $n\sim 2$ to $R_A(0.6)\sim =0.94$ which is  about 1/2 of the observed effect.
So this approach when  including the   Bjorken definition of $x$ and taking into account   the contribution of equivalent 
photons  has problems to describe observed dependence of $R_A$ on $x$ and absolute value of $R_A$.

\subsection{Scaling violation and Fermi motion effects}
\label{subsection 3.1}
The experimental data on $F_{2N}(x, Q^2)$ for large $x$  indicate that the x-dependence of $F_{2N}(x, Q^2)$ changes 
strongly with $Q^2$. Usually this effect is interpreted as due to the presence of the higher twist effects for small $W$ / low $Q^2$ which can be written in the  form 
\begin{equation}
F_{2N}(x, Q^2) = F_{2N}^{LT}(x, Q^2) (1+ {c\over  (1-x)Q^2}+ 
1+ {d\over  (1-x)^2Q^4}).
\label{HT}
 \end{equation}
The fits to the SLAC and Jlab data corresponding to typical $Q^2 = \mbox{5 GeV}^2$  give $F_{2N}(x, Q^2) \propto (1-x)^n, n=2$ \cite{lowQ}  while for the large x data taken at CERN  typical $Q^2=\mbox{40 GeV}^2$ , $n\sim 3$   \,\cite{Benvenuti:1989rh}.
This increase of the effective $n$ with $Q^2$
is consistent with the expectations of Eq. \ref{HT} that the HT effects should die out with increase of $Q^2$ for fixed $x$.

\begin{figure}
\begin{tabular}{ll}
\includegraphics[width=.40\textwidth]{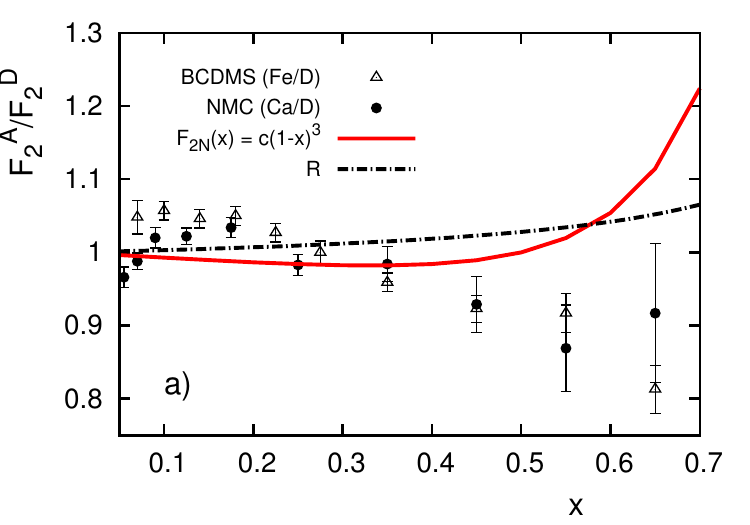} &
\includegraphics[width=.40\textwidth]{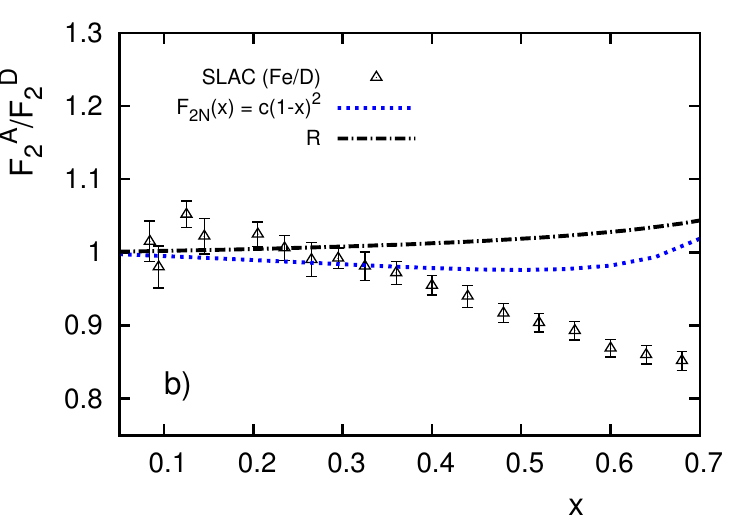}
\end{tabular}
\caption[]{Effect of Fermi motion on the EMC ratio for large and moderate $Q^2$ - solid and dashed curves and the effect of f proper definition of $x$ (with a small correction due to equivalent photons) - dot-sashed curved.   The data are from   \cite{Gomez:1993ri} -- right figure and from \cite{Benvenuti:1987az,Amaudruz:1991dj} - left figure.}
\label{medium}
\end{figure}

One can see from Eq. \ref{Fermimotion} that the Fermi motion effect is strongly modified when $n$   changes from 2 to 3 ,
cf.  Fig. \ref{medium}.  Also with increase of $x$ average  $x/\alpha$ in the convolution integral start to exceed significantly 
$x$ leading  to enhancement of the higher twist contribution to the EMC ratio.  The correction due to  the difference between 
$x$ and $x_p$ increases  between low  and high $Q^2$ by a factor of $\sim$ 3/2. Similarly the  effect of the extra degrees 
of freedom in Eq. \ref{Fermimotion} also changes by a factor of $\sim$ 3/2.

Note in passing that there may exist specific nuclear HT effects. One example is the  quasielastic contribution which 
becomes significant for moderate $Q^2$ for $x\ge 0.8$. Another potential source of the nuclear HT effects is scattering 
off the SRC where 6 quarks come rather close together.

\section{Photon distribution in  nucleons and  nuclei.}
\label{section4}
\subsection{General framework}

The long range Coulomb field of nucleus at rest is the zero component of the electromagnetic field.  Lorentz transformation 
of nucleus  Coulomb field from rest frame to the frame where nucleus is rapid unambiguously leads to the field of Fermi,  Weizsacker and Williams  equivalent photons \cite{WW}. Thus the  light-cone WF of nucleus contains photons as 
constituents carrying a fraction of the nucleus momentum.   Collision of equivalent photons of one nucleus with nucleon 
or  nucleus beams produces variety of hard and  soft   high energy processes in the  ultraperipheral processes at  the LHC \cite{Baltz:2007kq}.  Ultraperipheral  processes were observed at RHIC, see review in \cite{Baltz:2007kq}.  

For this paper a proper example of the ultraperipheral processes is  the diffractive production of massive lepton pairs 
$\gamma^{*}+A\to L^{+}+L^{-}+A$.  Large $Q^2$ and/or large mass of the  lepton pair  $L^+L^-$   guarantees 
dominance of the leading twist   term and  allows one to define photon distribution in a nucleus, $P_A(x,Q^2)$.
 
\begin{figure}[t]  
   \centering
   \includegraphics[width=0.5\textwidth]{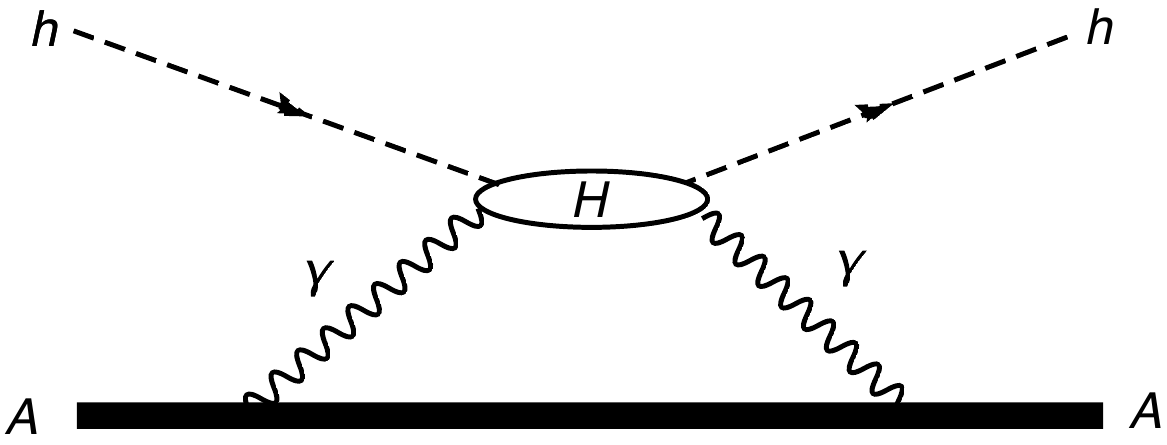} 
   \caption{Diagram for the interaction of photon of the nucleus with a hard probe $h$.}
   \label{plot}
 \end{figure}

A nucleus is characterized by quark, gluon, photon  distributions within a nucleus.   To suppress particle production by hard probe from vacuum   the gauge condition $A^{+}=0$ is chosen,  where $A_{\mu}$ is the operator of photon field.     In this gauge  photon distribution has  the form familiar from  QED,   cf.\, Fig.~\ref{plot}.   The photon parton distribution can be written as the matrix element of the product of the operators cf.  \cite{Sterman}:
\begin{eqnarray}
P_{A}(x,Q^2)={1\over \pi} (2\pi xp^{+})^{-1} \int_{-\infty}^{\infty}dy^{-} \exp{(-i(x/A)p^{+}y^{-})}\nonumber \\ \cdot 1/2 \sum_{\mu}\langle A\left|\left[F^{+}_{\mu}(0,y^{-},0),F^{\mu ,+}(0)\right]\right|A\rangle_{A^{+}=0}.
\label{parton1}
\end{eqnarray}
Here  $F_{\mu,+}$ is the operator of the strengths  of the photon field with transverse component $\mu$,    and $x/A=Q^2/2(p_Aq)$ is the fraction of nucleus momentum carried by a parton.

To simplify calculations it is convenient to represent $P_A$  as the sum of two contributions  $P_A=P_A^{inel}+P_A^{coherent}$.  First term accounts  for the  target excitations by virtual photon - we will refer to it as the  inelastic term. The second  term is  the contribution of equivalent photons.  We will refer to it as the coherent  term.

Small value of  $\alpha_{em}$  guarantees that  in the kinematic domain achievable at accelerators  
$\alpha_{em}\ln(Q^2/\lambda^{2}_{QCD})/2\pi \ll 1$ and therefore all effects of this order can be neglected. In 
particular, in this kinematic domain the effects of running of $\alpha_{em}(Q^2)$ due to combined QED and QCD effects  
are negligible.   Moreover it is legitimate to neglect in this kinematics by the $Q^2$ evolution of the amplitude of  the 
scattering of a hard probe  $h$ off the photon   which leads to   tiny    corrections.    There are two practical consequences.  
The $Q^2$ evolution of  inelastic
 photon distribution in a target $T$  -- $P_T(x,Q^2)$  goes in one direction: quarks, antiquarks radiate photons in the process of evolution but a photon does not radiate them. This evolution is accounted for in  the evolution equation through the $Q^2$ dependence of  valence, sea quark and gluon densities.  At $1/2m_N R_A\le x \le 0.6$  where  nuclear shadowing (anti shadowing) effects and nucleon short range correlations  are a small correction, the  account of inelastic photon distribution does not violate additivity and therefore does not change significantly $R_A$ -   (such an effect is  suppressed since it is proportional to a product of two small  factors: the  smallness of e.m.correction and the overall smallness of the EMC effect). 
   
  Second consequence of the  smallness of the electromagnetic constant is that coherent contribution is not evolving  with $Q^2$ at  $Q^2$ achievable experimentally. Coherent contribution due to $Z^2$ dependence violates additivity and changes $R_A$ .

Above we defined  $P^{inel}_A$  exactly in the same way as other parton densities,  cf. Eq.~\ref{parton1}.  This definition allows 
to calculate $P^{inel}_A$ in terms of quark, antiquark nuclear pdfs by evaluating  corresponding Feynman diagrams.
\begin{equation}
xP^{inel}_A(x,Q^2)/A={\alpha_{e.m.}\over \pi} \int_0^{k^2_{t max}} d k_t^2 
 \int _{\nu_{min}}^{\nu} {d\nu'\over \nu'} 
{k_t^2 \over (k_t^2+Q^2\nu'/\nu)^2} F_{2A}(\nu',k_t^2)/A.
\label{photon}
  \end{equation}
Within the leading $\alpha_{e.m.}\log k_t^2 $   
approximation upper limit of integration over photon transverse momentum 
is  ensured by the square of nuclear form factor . So $k^2_{t max}\approx (3/r_A^2) \le Q^2$.   
Since dominant contribution arises from small $k_t$  sensitivity to nuclear form factor at essential $x$ is weak.
Here $\nu=2(pq)/A=Q^2/x$,  
$k_t$ is the transverse momentum of the photon, and $F_{2A}(\nu,Q^2)/A$ is the virtual Compton amplitude normalized per 
nucleon.  In the above formulae we neglected   the small contribution of  the  longitudinally polarized  photons ($F_L^A$).  

The presence of  equivalent photons in the nuclear WF   leads to the violation of   the 
intuitive prediction we discussed in sections \ref{section2}, \ref{section3} -- 
$R^{j}_A(x,Q^2)=1$, for $x\le 0.5$ :
\begin{equation}
R^{j}_A(x,Q^2)= {Z f^{j}_p(x/(1-\eta_{\gamma}),Q^2) + N f^{j}_n(x/(1-\eta_{\gamma}),Q^2) \over  Z f^{j}_p(x,Q^2) + N f^{j}_n(x,Q^2)}.
\label{A/N}
\end{equation}
Here $\eta_{\gamma}(A)$ is the fraction of nucleus momentum carried by equivalent photons calculated in \cite{Frankfurt:2010cb}. 
Account of  the presence of  $\eta_{\gamma}(A)$  leads to a definite   dependence of $R^{j}_A$  on $Z$ and $A$. For 
example this effect is  as large as   the effect of accounting for the difference between $x$ and $x_p$ for heavy nuclei
but opposite sign  while it is negligible for Z=2.

\subsection{The coherent and incoherent contributions into photon distribution of a nucleus}

First we will calculate coherent contribution to parton nucleus distribution  which dominates the photon distribution in a nucleus. 
This contribution to $P_A(x,Q^2)$ arises from the interaction of a hard probe with a photon coherently emitted by   the target 
so that the target remains intact,  cf. Fig.\ref{plot}.   The coherent contribution to the photon structure function  is 
unambiguously calculable  in terms of the electromagnetic form factors of  the nucleus target.  For a proton target coherent
term has been calculated in \cite{Kniel}. Inelastic contribution for a nucleon target (where nucleon is excited in the final state)  was calculated in the perturbative QCD model where quarks and gluons are generated via evolution starting at very small $Q^2$ \cite{Gluck:1994vy}. Comparison of two contributions was performed at \cite{Gluck:2002fi} where it was found  that the elastic contribution is much more 
important in the proton case in a wide range of the $x,Q^2$. In the neutron case incoherent contribution is more important.

An important aspect of the photon contribution -- its implications for the momentum sum rule was not discussed in these papers 
and effects of the presence of photons in addition to quarks and gluons at the normalization scale
are  still not  included in the pdf studies by the groups which analyze the hard processes. We calculate the field of equivalent photons through the evaluation of Feynman diagram in Fig.\ref{plot}.  Calculations are simplified in our case since the  nucleus is heavy so the static approximation should be sufficiently accurate. In the static approximation  zero component of photon momentum in the nucleus rest frame is negligible:  $k_0=k^2/2m_A\approx 0$.  So
\begin{equation} 
 x=A(k_0-k_3)/M_A\approx -k_3/m_N.
\end{equation}

Second simplification arises from the observation that four-vector  $k$ can be decomposed over  directions defined by external momenta: 
$k_{\mu}=ap_{\mu}+bq_{\mu}+k_t$.    Here $p$ is four momentum of target nucleus and $q$ is four momentum of virtual photon (external 
hard probe) and $(p k_t)=(q k_t)=0$.  In the essential region:   
$p_{\mu}A_{\mu, \lambda}\approx (k_{t \mu}/a) A_{\mu,\lambda}$,  
cf. \cite{GribovComplex}. Account of this property leads to the generalization of the Fermi - Weizsacker -Williams expression 
for the spectrum of the equivalent photons:
\begin{equation}
xP^{coherent}_{A}(x,Q^2)={\alpha_{em}\over \pi } {Z^2\over A}
\int  d k_t^2 k_t^2  \frac{F_A^2(k_t^2+x^2 m_N^2)}{(k_t^2+x^2 m_N^2)^2},
\label{ww1}
\end{equation}
Here  $F_A(t)$ is the observed electric form factor of the nucleus which is equal to the product of the nucleus body form factor 
and  the proton electric   form factor $F_p(t)$.  Hence the characteristic  values of  $k_t^2$ are small.

In our  estimates we choose $F_A$ in the exponential form, i. e. 
\begin{equation}
 F_{A}(k_t^2+x^2 m_N^2)=\exp(-r_A^2 (k_t^2+x^2m_N^2)/6).
 \end{equation}
\noindent
Here $r_A$ is the experimentally measured  RMS nuclear radius and $m_N$ is the  nucleon mass.
 Such a form  allows one  to perform the  numerical
 integration over  the transverse momenta of the photons.

It is useful to compare different electromagnetic contributions.  The most important one is the contribution of equivalent photons 
in which  the fields of individual protons add coherently.   This coherence leads to a larger momentum fraction carried by the 
photon field in nuclei as compared to that carried by  individual free protons.  Eq.~\ref{ww1} can be used to evaluate   the 
coherent contribution of the photons to the momentum sum rule (here we already subtracted contribution of individual protons to 
the photon parton density - see discussion below, since this quantity enters into description of the nuclear effects we discuss in 
the paper):
\begin{equation}
 \eta_{\gamma}=\int_0^1\, dx xP_A(x,Q^2)={\alpha_{em}} {2\over \sqrt{3\pi}} {Z^2\over A}{1 \over m_Nr_A}.
  \label{ww5}
 \end{equation}
 \noindent
 
To separate nuclear effects one need to  subtract the contribution of Coulomb fields of protons. This is achieved by taking
 into account contribution of the incoherent break up of the nucleus as well as production of hadrons (inelastic processes).  Sum 
 of the two effects can be calculated in the closure approximation where cross section is described by the sum of two diagrams 
 presented in   \cite{Frankfurt:2000jm} and   results in  replacing $ Z^2F_A^2(t) \longrightarrow  ZF_N^2(t) + Z(Z-1)F_A^2(t)$ in 
 Eq. \ref{ww1} leading to 
    \begin{equation}
 \eta_{\gamma}=\int_0^1\, dx xP_{A}(x,Q^2)={\alpha_{em}} {2\over \sqrt{3\pi}} {Z(Z-1)\over A}{1 \over m_Nr_A} + {Z\over A}\eta_p
  \label{replace}
 \end{equation}
 \noindent

The  second  term in Eq.\ref{replace} is due to the Coulomb field of the individual protons and it is included in the structure functions of the proton. Hence it should not be included in the calculation of the nuclear effects. The contribution of magnetic form factors of neutron and proton is negligible because it is concentrated at  larger momentum transfers  than the scale of the  nuclear phenomena.

Taking $r_A$ from the compilation of ref.\cite{Angeli} we find 
  for the nuclear term in Eq.\ref{replace}
  \begin{eqnarray}
\eta_{\gamma}(^4He) = .03\%; \eta_{\gamma}(^{12}C) = .11\% ;\eta_{\gamma}(^{27}Al) = .21\% ;\nonumber \\
 \eta_{\gamma}(^{56}Fe) =.35\%;
\, 
\eta_{\gamma}(^{197}Au) =.65\%.
\end{eqnarray}
To evaluate the impact of the presence of the photon  component for the nuclear  structure functions we can use 
Eq.\ref{Fermimotion} and take $\eta_A=\eta_{\gamma}(A)$.  The small $x$ nuclear shadowing effects modify this 
approximation, however they are negligible for $x\ge 0.2$ range we are interested in
and give insignificant contribution into momentum sum rule. 
Since  deviations from the 
additivity are small the effect of the presence of the photons and other  "hadronic" effects can be treated as 
contributing additively to the deviation of the EMC ratio  from one.  Larger numbers given in \cite{Frankfurt:2010cb} 
are because of misprint in the formulae for coherent contribution and due to inaccurate numerical calculation.

\subsection{Equivalent photons and modification of the  initial condition for the  $Q^2$ evolution of nucleus and  nucleon pdfs }

Major impact of  the presence of the photon constituents in a nucleus is through  the change of  the form of the energy-momentum conservation and violation of isotopic invariance in the incoherent contribution of protons and neutrons.
Total fraction of nucleus momentum carried by QCD partons instead of 1 becomes:   $1-\eta_{\gamma}(A)$ (in this fraction 
the photon fields due to individual nucleons are included).

The presence of the photon component in the nuclear light-cone WF  leads to the certain modification of $Q^2$ evolution of parton densities. Most  important is the change  of  the  momentum sum rule due to necessity to account for the fraction of nucleus momentum carried by photons:
 \begin{eqnarray}
 \int_0^A \left[(1/A)(xV_{A}(x, Q^2) +xS_{A}(x, Q^2) +xG_{A}(x,Q^2))\right]dx=\nonumber \\ 1  - \int_0^A \left[(1/A) xP_{A}(x, Q^2)\right]dx=1-\eta_{\gamma}(A). & &
 \label{sumrule}
 \end{eqnarray}
The  light cone momentum carried by equivalent photons is compensated by  the loss of the momentum by the nucleons.  
Since the  nucleus  Coulomb field is the collective field dominated by the contribution of large impact parameters, the 
reduction of the light cone fraction is experienced by nucleus as a whole (i.e. by both protons and neutrons.) Hence the 
shift is approximately equal for protons and neutrons.

Although $\eta_{\gamma}(A)$ is small  its role is enhanced by the rapid decrease of nuclear pdfs with $x$ increase.

In the case of a nucleon target one should also account for the change of the form of the  momentum conservation 
and much larger contribution of photons for the proton target than for the neutron target leading to an isospin violating effect for pdfs of protons and neutrons and hence for the nuclear targets.

\section{Implications for  the EMC effect  for nuclear pdfs at $x\le 0.5$}
\label{section5}
Here we study to what extent  two effects we discussed above - the proper definition of  $x$ which preserves 
the momentum  sum rule, and the  account of the collective Coulomb field of the nucleus (of the momentum carried by 
equivalent photons) - which we refer below as the standard model of the nucleus - explain the data at $x\le 0.5$ 
where Fermi motion is a small correction.

 In the standard model we can use  Eqs.  \ref{Fermimotion},\ref{rs}  to obtain
\begin{equation}
R_A^j(x_{p})= f^j_A(x(1+r_x+\eta_{\gamma}))/f^j_N(x)=1  -(r_s+\eta_{\gamma})n{x\over 1-x}+ {n x  (x(n+1) -2) \over (1-x)^2}\cdot 
{T_A\over 3m_N}.
\label{rstot}
\end{equation}
Since the EMC ratio is experimentally defined relative to the deuteron one needs to substitute $r_x(A) $ by $r_x(D)$.   
The results of calculations using Eq.\ref{rstot} with $T_A$ from \cite{CFKS} are presented in Fig. 2 for  medium A nuclei 
for low and high $Q^2$.  We show separately the effect of Fermi motion and the combined effect of the account for the 
x-scale  and for Coulomb effects.

One can see that the the Fermi motion effect is very different for low and high $Q^2$ especially for $x \ge 0.55$.
  \footnote{The data for heavy nuclei \cite{Gomez:1993ri}  were corrected for the difference of the number of protons and neutrons. 
}.

Overall inspection of Fig.\ref{medium} indicates that the  contribution
to EMC effect due to the modification of the quark distribution in nucleons is significant only for $ x \ge 0.5$.

The effect of the correcting for the difference of $x_p$ and $x$
leads to an increase of the EMC effect which is very similar to all nuclei with $A\ge 4$ since the energy binding is a weak function of A. Effect is much smaller for the case of  $^3$He  namely for the ratio  $F_{2^3He}(x_{p})/[F_{2\, ^2H}(x_{p})+F_{2p}(x_{p})]$ due to a very small energy binding per nucleon in $^3$He. Absolute value of the EMC effect for this ratio is rather uncertain since the magnitude of the effect is comparable to the normalization uncertainty \cite{Daniel:2011ng}.  However an effect on the level of 1/4 of the effect for heavy nuclei (the A-dependence  expected from the contribution of the short-range correlations - see discussion below)  -  cannot be excluded.).
 
 Overall the proper choice of the  x-scale --
 that is the choice of Bjorken $x$ --
   increases the EMC effect
 by $\sim 15\div 20 \%$ for $^4$He and other light nuclei and a factor of two smaller  for heavy nuclei.
 
 At the same time as we have already discussed in the case of the heavy nuclei the Coulomb effect contributes to 
 the increase of the EMC effect reducing the  hadronic component of the EMC effect. 
 For $A\sim 200$ two effects practically compensate each other (see Fig. ~1). 
 
Overall the discussed effects lead to  a larger EMC effect for light nuclei and a weaker A-dependence of the 
hadronic component of the EMC effect  for $A \ge 4$.

\section{Recent progress in the studies of  SRCs in nuclei}
\label{section6}
Before discussing the A-dependence of the the hadronic EMC effect and its dynamical origin we need to discuss 
the information about the short-range correlations in nuclei.   A very significant  progress in this field has been 
reached in the last few years. As a result it has  become possible to perform nearly   model  independent  tests of 
the role of the SRC in the EMC effect and also strongly 
constrain the   models of the  EMC effect.

Hard interactions resolve local structure of the nucleus. Hence  they resolve collective low energy scale degrees of freedom (effective interactions) and probe local structure of the nucleus. 
Since the nuclear forces are short range on the nucleus scale ($r_{NN} < 2 fm$)  the properties of a nucleon in the nuclear media are determined by its local surrounding. Intuitively it is clear that  closer the nucleons get together, the  stronger their
 polarization/deformation  is. Hence the configurations where nucleons nearly overlap appear to be a natural candidate for the large hadronic EMC effect.  Therefore  we need to review briefly the recent progress in the studies of SRCs.
 
Singular nature of the  $NN$ interaction in coordinate space at  small inter nucleon distances/large momenta leads to the universal structure of SRCs and  to the  prediction of the scaling of the ratios of the cross sections of $x  > 1$ scattering at sufficiently large $Q^2 \ge 2 \mbox{GeV}^2$ \cite{Frankfurt:1981mk}. In particular for $1+k_F/m_N < x < 2$: 
\begin{equation}
R_A(x,Q^2)={2\over A} {\sigma(eA \to e + X) \over \sigma(e^2H \to e + X)} =  a_2(A).
\label{a2ratio}
\end{equation}
Here $a_2(A)$ has the  meaning of the relative probability of the two nucleon SRCs per nucleon in a nucleus and in the deuteron, 
that is, it is   the ratio  of spectral functions of nucleus to that of deuteron.
Actually it includes the contribution of triple nucleon correlation, pn and pp correlations.

The first evidence for such scaling of the ratios was reported in \cite{Frankfurt:1988nt}.    The extensive studies were 
performed in \cite{Frankfurt:1993sp} using various data taken at SLAC  at  somewhat different settings  which confirmed 
the scaling of the ratios and for the first time confirmed the prediction \cite{Frankfurt:1988nt} of the  "super" scaling of the 
ratios at different $Q^2$   -- the precocious scaling of the ratios plotted as a function of $\alpha_{t.n.}$ -- the minimal $\alpha$ for the scattering off two nucleon SRC at rest (the Fermi motion of the pair practically cancels out in such a ratio)
\cite{Frankfurt:1988nt}: 
\begin{equation}
\alpha_{t.n.}= 2 -{q_0-q_3+ 2m_N\over  2 m_N}\left(1+ {\sqrt{W^2-4m_N^2}\over  W}\right),
\end{equation}
 where $W^2=4m_N^2+4 q_0m_N-Q^2$.

  The experiments performed at JLab allowed to explore the scaling of ratios in the same experiment. 
In \cite{Egiyan:2003vg,Egiyan:2005hs} the scaling relative to $^3$He was established. Very recently the results of the extensive study of the  nucleus/deuteron ratios were reported in \cite{Fomin:2011ng} allowing a high precision determination of the relative probability of the two nucleon SRCs in nuclei and the deuteron.  The results of \cite{Fomin:2011ng} are in a good agreement with the early analysis of \cite{Frankfurt:1993sp}, see Fig.~3.

Several theoretical observations are important for the interpretation of the scaling ratios: (a) The invariant energy of 
the produced system for the interaction off the deuteron is large on 
the scale of nuclear phenomena but small as compared to the scale characteristic  of  hadronic phenomena:
 $W - m_{^2H} \le \mbox{250 MeV}$. This kinematics obviously leads to a strong suppression of the production of the inelastic final states.  Correspondingly, scattering off  exotic configurations  which decay into excited baryon states,  $\Delta$'s, etc is strongly suppressed in the discussed kinematics. (b) The closure is valid for the final state interaction of  the nucleons of the SRC and the residual nucleus system. Only the f.s.i. between the nucleons of the SRC contributes to the total (e,e') cross section \cite{Frankfurt:1993sp,Frankfurt:2008zv}. Since this interaction is the same for light and heavy nuclei it does not modify the scaling of the ratios. (c) In the limit of large $Q^2$ and large energy $q_0$  but the fixed ratio 
$x=AQ^2/2m_{A}q_0$,  the cross section is expressed through the light - cone projection of the ground state
nuclear density matrix, $\rho_A^N(\alpha)$, that is the integral over all components of the interacting nucleon four momentum except $\alpha\equiv p_-/(m_A/A)$, where $p_- =p_0-(\vec{p}\cdot \vec{q})/ |\vec{q}|$. The ratio of the cross sections reaches a plateau at $x(Q^2)$ corresponding to the scattering off a nucleon with minimal momentum $\sim k_F$ indicating that the dominance of two nucleon SRC sets in just above the Fermi surface. (d) Observation of the precocious  $\alpha_{t.n.} $ scaling indicates that $R_A$ is equal to  the ratio of the light-cone density matrices of the  nucleus and deuteron. It also strongly indicates that  SRCs of the baryon charge two are predominantly built of two nucleons rather than some exotic states. It is worth noting here that the correspondence between the light cone and nonrelativistic wave functions is pretty straightforward. Hence the results  for the ratio of wave functions in the region where pair correlations dominate should be  close. 
\begin{figure}
  \includegraphics[height=0.4\textheight]{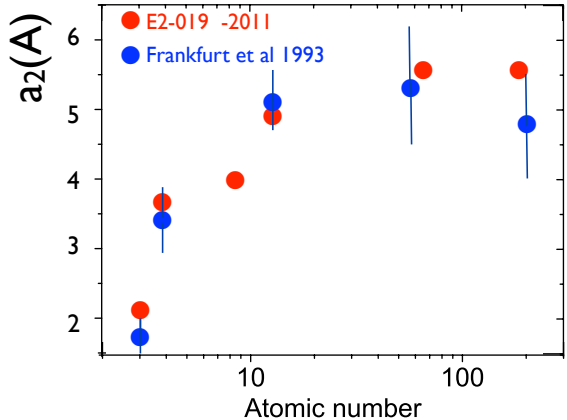}
\caption[]{Comparison of the first determination of $a_2(A)$ based on the analysis of the SLAC data  \cite{Frankfurt:1993sp}with the most recent Jlab measurements \cite{Fomin:2011ng}.   }
\end{figure}

To probe directly the structure  of the SRCs it is advantageous to study a decay  of SRC after one nucleon of the SRC is removed which is described by the nuclear decay function
 \cite{Frankfurt:1981mk,Frankfurt:1988nt}. In the two-nucleon SRC approximation the decay function is simply expressed through the density matrix as the removal of one of the nucleons of the correlation results in the release of the second nucleon with probability of one.  A series of the experiments was performed at BNL and JLab which studied (p,2p), (e,e'p)  reactions in the kinematics where a fast proton of the nucleus is knocked out (see review and references in \cite{Subedi:2008zz,Frankfurt:2008zv}). In spite of very different kinematics -- removal of forward moving nucleon in the $^{12}$C(p,2p) case and backward moving proton in the $^{12}$C(e,e'p) case, different probes and different momentum transfer $-t \approx \mbox{5 GeV}^2 $ and $Q^2=\mbox{2 GeV}^2$ -- the same pattern  of the neutron emission was observed:  the neutron is emitted with a probability $\sim 90\%$ in the direction approximately opposite to the initial proton direction with the correlation setting in very close to $k_F(C) \sim $ 220 MeV/c. The JLab experiment  observed in the same kinematics both proton and neutron emission in coincidence 
with $e'p $ and found the probability of the proton emission to be about 1/9 of the neutron probability. Hence the data confirm  our theoretical expectation that removal of a fast nucleon is practically always associated with the emission of the nucleon in the opposite direction with the SRC contribution providing the dominant component of the nuclear WF starting close to the Fermi momentum. The large pn/pp ratio also confirms  the standard expectation of the nuclear physics that short-range interactions are much stronger in the isospin zero channel than in the isospin one channel and hence are much more sensitive to the pion-like exchange. 
Saturation of the probability provides an independent confirmation of the conclusion that at least up to momenta $ \sim 500 \div 600 $ MeV/c, SRC predominantly consist of two nucleons. 

A word of caution is necessary here. In the standard nuclear physics approach the intermediate $\Delta$ isobars play an important role in the $I=1$  interaction channel. Usually they are absorbed into the definition of the low energy $NN$ potential. However  in the case of hard exclusive process the $N\Delta$ SRCs are resolved and may show up and be comparable to $pp$ SRCs. 

The structure of SRC depends on $x,Q^2$ and $M^2$. Here $M^2$  is the mass range allowed for the decay of SRC. For $M$
close to $2m_N$  possible exotic states enter into parameters of effective low energy Hamiltonian.  With increase of $M$ exotic states may reveal themselves in the decay of SRC. Thus $M - 2m_N$ plays the   role of the  resolution and increasing resolution allows one
to investigate some properties of superdense matter which is present  in the inner core of neutron stars. Note here that the allowed phase volume was much larger in the BNL   A(p,2pn) experiment. So the consistency of two experiments suggests that 
 increase of the resolution does not lead to a drastic change in  the structure of the short range nuclear structure.

\section{Short-range correlations in nuclei and  the A-dependence of the hadronic EMC effect}
\label{section7}

We will show below that the strength of the deformation of the bound nucleon WF as compared to the free nucleon WF  
is proportional to its  off-shellness. At $x< 0.7$ where the Fermi motion effects are small it is possible to average over nucleon momenta and conclude that the deformation of the structure function of a bound nucleon  should be proportional to the nucleon average kinetic energy $T_A$.  Since $T_A$ is dominated by the contribution of the short range correlations \cite{CFKS} in 
nuclei we can roughly estimate the A-dependence of $T_A$ from the measurements the ratio of the high momentum components 
in the nuclei and in the deuteron - $a_2(A)$. This ratio is equal to the ratio of the nucleus and the deuteron inclusive (e,e')  cross sections in the x region  of $1.4 < x < 1.8$ where it exhibits a $Q^2$ independent plateau - (see  discussion in Section \ref{section6} and in particular Eq.\ref{a2ratio}).

 This information can be used to predict the A-dependence of the hadronic EMC effect for $x=0.5 \div 0.7 $ where nucleon Fermi motion is  small for $F_{2N}(x,Q^2) \propto (1-x)^n, n=2$ corresponding to the SLAC/Jlab kinematics (cf. Fig.\ref{medium}). Since the Fermi motion is a few \% correction which is proportional to the average $T_A$ and most of the kinetic energy originates from the SRCs the overall A-dependence of the hadronic EMC effect including Fermi motion effect should be approximately proportional to $a_2(A) -1$.
 
\begin{figure}
\includegraphics[width=0.5\textwidth]{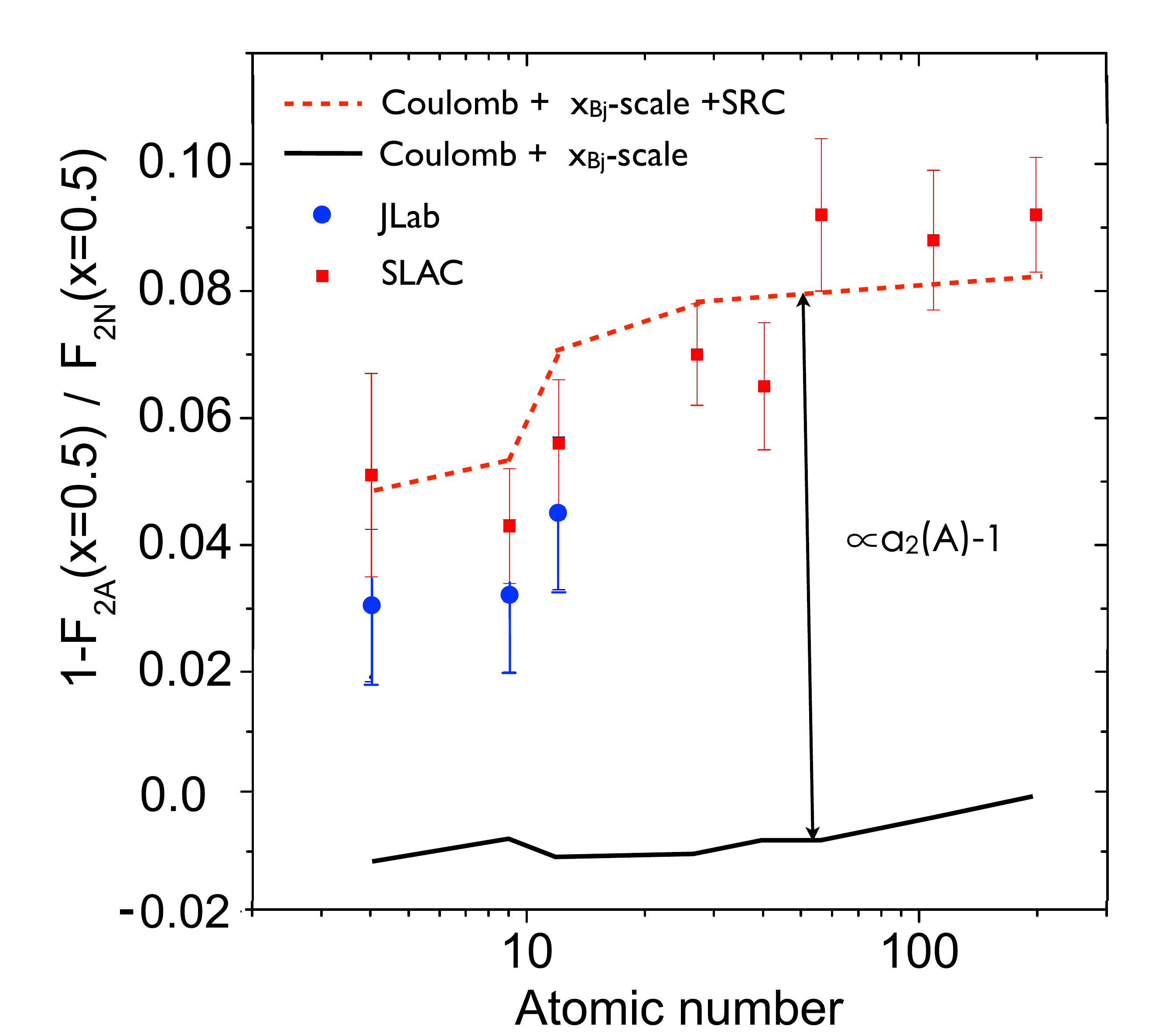} 
  \caption[]{The solid line is the  result of calculation taking into account the   equivalent photon effect
 and the effect of proper definition of $x$. The dashed line is the contribution of the hadronic  EMC effect due to SRCs normalized for large A with the  A-dependence $\propto  a_2(A)-1$ with  $a_2(A)$ from  taken from the JLab measurement \cite{Fomin:2011ng}. The ratio data are from 
 \cite{Gomez:1993ri,Seely:2009gt}.
 }
    \label{x05}
 \end{figure}

 One can see from Figs. \ref{x05} that the A-dependence of the "hadronic" EMC effect for  $x=0.5$ is indeed consistent with the A-dependence of $a_2(A)-1$. It is worth noting here that in \cite{Frankfurt:1985cv,Frankfurt:1988nt} it was assumed that {\bf all} $x\ge 0.4$ EMC effect is due to the contribution of the SRCs.  (Note here that nucleons belonging to mean field give up to 20\% contribution to $T_A$, leading to a deviation of the A-dependence of $T_A$ and $a_2(A)$.)

\begin{figure}
\includegraphics[width=1.0\textwidth]{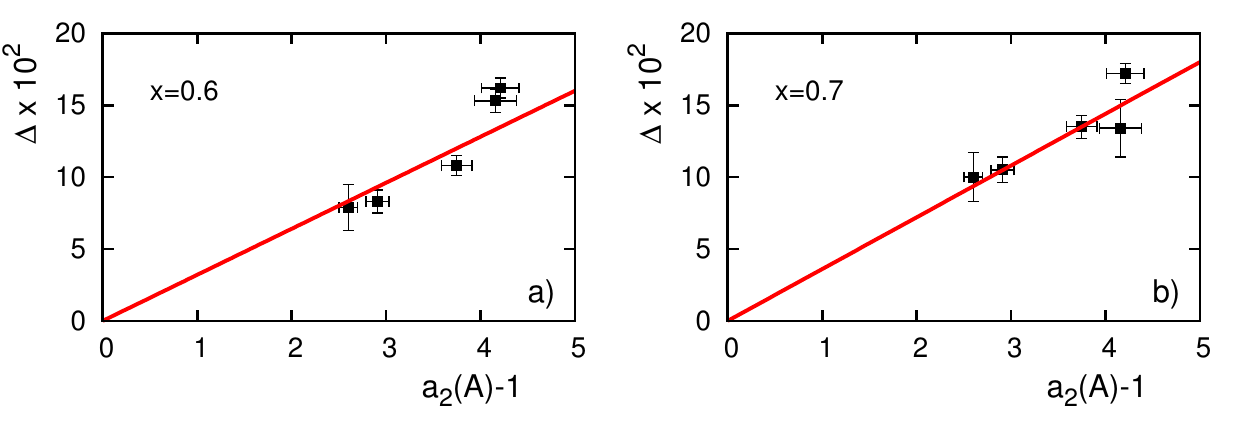} 
  \caption[]{A-dependence of the hadronic EMC effect for  $x=0.6 $ and $x=0.7$. The data are from \cite{Gomez:1993ri}. }
    \label{x0607}
 \end{figure}

  The data at $x=0.6$ and $x=0.7$ where the Fermi motion effect is practically zero for the SLAC kinematics are also consistent with hadronic EMC effect been proportional to $a_2(A)-1$. To illustrate this proportionality we calculate the   difference  between the standard model without Fermi motion and the data, which we denote as $\Delta(A)$ and plot it as a function of $a_2(A)-1$ --  Fig. \ref{x0607}.
  The data are consistent with $\Delta(A)\propto a_2(A)-1$ though more accurate data would be highly desirable. 
  
To analyze the A-dependence of the hadronic EMC effect at $x > 0.5$ we need to take into account the nucleon Fermi motion effect which rapidly increases with $x$ for $x\ge 0.55$ and which is $\propto T_A$ in the discussed $x$-range (Eq.\ref{sercon}) and results including Fermi motion --  dashed curves in Fig. 2a, 2b).  One can see that the difference between dashed curves and the data is much larger than the difference between the standard model without Fermi motion and the data already for $x$=0.6. This implies right away that the hadronic EMC effect is roughly  $\propto T_A$. To check it with a better accuracy we can focus on the A-dependence of the difference  between the standard model without Fermi motion and the data, which we denote as $\Delta(A)$. In Fig. \ref{x0607} it is plotted as a function of $a_2(A)-1$. The data are consistent with $\Delta(A)\propto a_2(A)-1$ though more accurate data would be highly desirable. Thus we conclude that in the range: $0.5  \le  x \le 0.7$ the hadronic EMC effect is approximately proportional to the probability of two nucleon correlations in nuclei or practically equivalently to the  average nucleon kinetic energy of nucleon in nucleus.

 For larger $x$ this pattern should break down since the expansion in powers of (1-$x$) breaks down. In fact for $x\sim 1$ we expect $F_{2A}/F_{2\, ^2H} \propto a_2(A)$ rather than  $F_{2A}/F_{2 \,{^2H}} -1  \propto a_2(A)$.

Theoretical and phenomenological arguments in favor of the similarity of the A-dependence of the EMC effect and the two nucleon SRCs were first presented in \cite{Frankfurt:1985ui,Frankfurt:1988nt} for $x\ge 0.5$.  New data on the A-dependence 
of SRCs allowed to demonstrate 
proportionality of the EMC effect to the probability of the two nucleon correlations in nuclei  without employing model for the 
A-dependence of SRCs \cite{Weinstein:2010rt}. The authors assumed that the SRCs dominate the EMC effect for the 
$R_A(x_p)$ ratio for all $x$.  They focused on the x-slope of the ratio in the $x=0. 35 - 0.7$ range which is less sensitive 
to the absolute normalization of the data assuming that the data can be fitted as a linear function of $x$.   As we have seen above  accounting for the standard model effects
(which in most of the  range studied in \cite{Weinstein:2010rt}   constitutes less than 20\% of the  difference of  $R_A(x_p)$ 
from unity)   does not change this conclusion qualitatively. Still it may suggest that the  A-dependence of the hadronic component of the  EMC effect between $A=4$ and $A=200$ is somewhat stronger than given  purely by SRCs.  
 
 As we discussed already in section 3 the Fermi motion effect  is very different for moderate  and  large $Q^2$.
 Assuming that the hadronic EMC effect can be described by the deformation of the bound nucleon wave function which is to the first approximation proportional to $T_A$ we can estimate
 deviation of 
 $R_{b} (x,Q^2) = F_{2N}^{bound}(x,Q^2)/F_{2N}^{free}(x,Q^2)$ from unity assuming factorization of the correction to the EMC ratio as a product of the nucleon modification effect and the Fermi motion and correcting for the x-scale and Coulomb effects. For $n=2$  and $x > 0.7$ 
 ($n=3$ and $x> 0.6 $ this procedure breaks down  since in the convolution larger than average nucleon momenta contribute
  leading to overestimate of $1- R_{b} (x,Q^2)$  for such $x$.  A separate analysis is required for this kinematics  which will be presented elsewhere.
 
\begin{figure}[t]  
   \centering
   \includegraphics[width=0.6\textwidth]{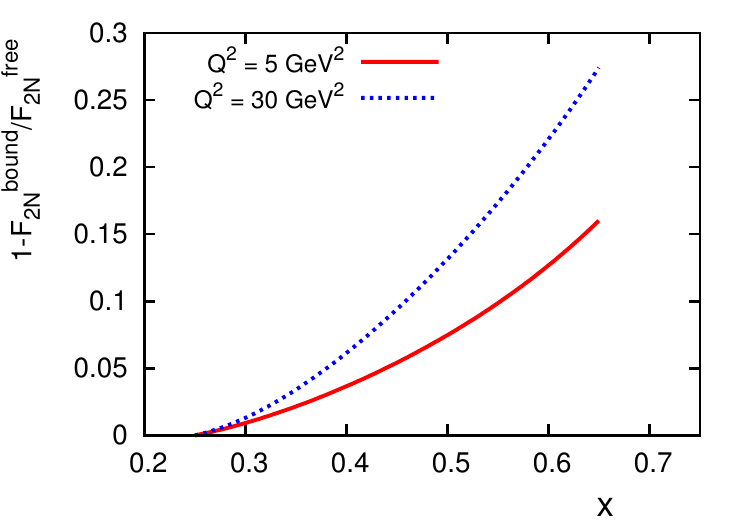}]
   \caption{Estimate of the ratio of the bound and free nucleon structure functions in medium and heavy nuclei as a function of $x$.}
   \label{bound}
 \end{figure}

  The estimate of $1 - R_{b} (x,Q^2) $  is presented in Fig. \ref{bound} based on the analysis  of the SLAC data\cite{Gomez:1993ri} for moderate $Q^2$ and BCDMS \cite{Benvenuti:1987az} and NMC  \cite{Amaudruz:1991dj} data for large $Q^2$.

We see that deformation rapidly grows with increase of $x$ starting at $x\sim 0.5$. There is a trend for the higher $Q^2$ data to indicate a stronger deformation of the bound nucleon wave function which may indicate a difference in the leading and higher twist effects in the interaction with the bound nucleon. However the errors of the high $Q^2$ data are very large for $x\ge 0.55$, see Fig.~2.

Since the bound nucleon deformation effect is small up to $x\sim 0.5$ and the probability of having a quark in a nucleon with $x > 0.5$ is $\sim 2\cdot 10^{-2}$  we conclude that the EMC effect probes very rare deformations  of bound nucleons. The deviations from the discussed approximation on the level of $ 2 -- 3 \%$ at $x\sim 0.2 \div 0.4$ cannot be excluded experimentally so one can only conclude that the accuracy of the approximation where nucleus consists of nucleons whose  structure  function coincides with that for free ones (+ equivalent photons)  is on the level of few \% for   $x\le 0.5$ where 
the contribution of the mean field approximation  to the nucleus WF dominates.

Our analysis also allows to put an upper limit 
\begin{equation}
 \eta_{\pi}(A\sim 200) \le 5.0\%.
 \end{equation}
on the fraction of energy carried by pion constituents of nuclei  assuming that all deviation of $R_A(x\sim 0.6, Q^2\sim 5 \mbox{GeV}^2)$ from the standard model is due to the pion field.    
 
\section{Constraints on the models of the EMC effect}
 \label{section8}
 
 In this section we   briefly summarize the constraints on the possible mechanism of the hadronic EMC effect
 and  confront different classes of models of the EMC effect   with these constraints. Majority of models  ignore 
 the effects described by the standard model and hence  should be modified  to account for the QCD dynamics
 to describe data. So we restrict our discussion by consideration of constraints on the
basic features of these models. The most important of these constraints are obtained from the following:
 \begin{itemize}
\item Additivity of  nucleon structure functions as a function of Bjorken $x$ follows from QCD dynamics of hard 
processes and parton structure of nucleons and nuclei 
to the extent that non-nucleonic degrees of freedom can be ignore  leading to
$R_A(x,Q^2)$  close to one.
\item  Probability conservation: the baryon and momentum sum rules.
\item  Difference between the fraction of nucleus momentum and nucleon carried by equivalent photons is
accurately calculated in QED.
\item The  hadronic EMC effect is  small for $x < 0.5$ and for larger $x$  rapidly increases with 
increasing $x$, see Fig.~7.
\item The A-dependence of the hadronic EMC effect is rather close to the A-dependence of the the short-range 
internucleon correlations, cf. discussion in section 7.
\item  High probability of short-range correlations $ \sim 20 \div 25\%$ and dominance of the nucleonic degrees of 
freedom ($ >$ 80\%) in the correlations, cf. brief review of existing data in section 6.
\item  No  enhancement of the antiquark distribution in nuclei is present for $x \sim 0.1$ 
with accuracy $\approx 1\%$ \cite{DY}.  
\item  $Q^2$-dependence of the magnetic  form factor  of  bound nucleons
with small momenta is very close to that of free nucleons \cite{Sick}.
\end{itemize}

Seemingly natural treatment  of the EMC effect   within the standard  nuclear physics approaches is to  explicitly 
include  mesonic degrees of freedom  in the nuclear wave function. Pions with small momenta 
$\approx m_{\pi}$ relevant for the  Chiral QCD Lagrangian  give insignificant contribution into antiquark distribution at 
$x\ge 0.2$. However  a   significant enhancement of the pion field of nuclei  at large pion momenta as compared to the 
system of free nucleons was found for example in \cite{Friman:1983rt}.  Within the nonrelativistic nuclear theory where
inner motion is unrelated to the c.m.  motion this prediction may be translated into  light cone dynamics where the  fraction of 
nucleus momentum (scaled by 1/A) carried by a pion is given by the formulae: $k_3/m_N$  where $k_3$ is the projection of pion momentum on 
the direction supplied by photon momentum. Prediction  of the enhancement of pion field at large pion momenta is 
in variance with the lack of modification of the antiquark distribution in the nuclei observed in the measurement of the Drell Yan pairs \cite{DY}.
To fit the data on the EMC effect authors  need to assume that mesons carry $\eta_{\pi} \sim 4 \%$ fraction of the nucleus
light cone momentum which is hardly consistent with  experimental restrictions on antiquark distribution in nuclei.   
The models which describe nucleus wave function by a vertex function with interacting nucleon off mass shell and the  
residual system on mass shell, are effectively in the same category as they have to compensate the violation of the 
momentum sum rule by introducing an additional (presumably mesonic) component in the nucleus wave function. These 
models typically predict an enhancement of the antiquark  $\bar u + \bar d$ distribution in nuclei of the order  of 10--20\% 
at $x\sim 0.1$ , while according to the data \cite{DY}  $\bar{q}_A/\bar{q}_N \le 1$ for this  $x$-range (the absolute accuracy 
of these measurements is of the order of 1\% ).   

Another class  of nuclear  models is  mean field models where a nucleon is treated as moving in the self consistent field 
of other nucleons with a deformation of a bound nucleon (nucleon swelling, ...)  independent from  nucleon momentum 
(for a review and references see \cite{Cloet:2010zz}).   These models ignore  the experimental observation of SRCs in 
nuclei,    and they do not provide an explanation for  the similarity of   the A-dependence of the hadronic EMC effect  to 
the A-dependence of the two nucleon SRCs. The similarity  of the $Q^2$-dependence of the nucleon magnetic form 
factors of the low momentum bound nucleons to that of the free nucleons is not  reconciled in these models with the  large hadronic EMC effect.  The fast onset  of the hadronic EMC  
effect with $x$ for  $x > 0.5$ and lack of the significant  effect at smaller $x$ were not predicted in these models. 
 
In a number of models it was assumed that nonnucleonic configurations  -- six quark configurations, 
$\Delta$ isobars, etc -- are present in nuclei with significant probability.  If the SRCs were an {\it incoherent} superposition 
of nucleonic and   non-nucleonic configurations one would not be able to generate an EMC effect larger than 20\% of the probability of the SRCs, which  is $\le 5\%$.   This is clearly insufficient to  explain the strength of the EMC effect of the 
order of 15\% for  $x\ge 0.6$  and $A \ge 40$.  If one  would assume that 
there exist both nucleonic SRC and exotic configurations with a  probability $\ge 20 \%$, necessary to fit the EMC effect,
the number of nucleons below the Fermi surface would drop below 60\% which is hardly consistent with the current experience of the nuclear physics.

 To summarize, the common  perception that there exist plenty of successful models of the EMC effect mostly 
arises from treating  the EMC effect as  an isolated phenomenon ignoring the totality of the constraints which were 
obtained from the studies of the hard nuclear phenomena in the past 20 years. At  the  moment there seems to be no 
viable alternative to the scenarios  where the EMC effect is associated with modifications  of rare  quark-gluon 
configurations selected by hard probe, and  which become larger with increase of 
 the bound nucleon momentum.

  \section{How nuclear medium modifies nucleon breathing }
    \label{section9}

\subsection {Deformations of bound nucleon wave function within QCD}

A solution proposed in \cite{Frankfurt:1985cv} is the mechanism 
where hard scattering off quarks selects at $x\ge 0.5$  rare quark-gluon  configurations in bound nucleons  
and that the  deformation of the bound nucleon WF  is enhanced for such rare configurations.
The mechanism is based on  two fundamental and well established  properties of QCD.

{\it Any characteristics of a composite system should fluctuate and depend on the process.}  The textbook example in 
QED (the abelian gauge theory) is the hydrogen atom where the calculation of moments of the hydrogen radius finds that 
$\langle<r^n\rangle= \int d^3r r^n \psi^2(r)$ differs from $\langle r\rangle^n$.  An example of 
observed  fluctuations in QCD is the
significant difference between the electric and axial radii of the proton:
$\langle r^2_{e.m.}\rangle^{1/2}\mbox{ =0.85 fm}, \langle r^2_{axial}\rangle ^{1/2}  \approx \mbox{0.65 fm}$. The fluctuations 
of strength of interaction 
also play a key role in the explanation of the phenomenon of high energy inelastic diffraction.
One can prove that in QCD  as in QED  interaction of a hadron in a  small size configuration with a hadron target  is much 
weaker than in an average  configuration. 
 
{\it 
Two patterns of fluctuations of interactions are well understood in QCD. } 
 One type of fluctuations  arises as the consequence of the dependence of interaction on the spatial size of color neutral 
 configuration.  This pattern is observed in particular in the color transparency phenomenon, for the recent review see  
 
 reveals in pion-nucleus interaction in the form of color transparency phenomenon in coherent production of two jets ,  cf.review and 
 corresponding references in \cite{Dutta:2012ii}. 
 
 Another pattern of fluctuations of strengths of interaction follows from the  dependence of interaction on the representation 
 of color group of $SU_c(3)$  characterizing constituents in color space.  This property  is 
 well known in respect to the invariant charge  as the dependence of Casimir operators of color group $SU_c(3)$ on representation:the ratio of Casimir operators for octet and triplet representations: $F^2(8)/F^2(3)=9/4$.   These fluctuations are
important for hard processes which we  discuss   in the next subsection. In the low energy processes instant interaction  is averaged out and cannot reveal itself.  On the contrary it follows from  the QCD factorization theorem that a hard probe selects particular instantaneous quark-gluon configuration in a hadron target.     Another feature of QCD is that the number of constituents is decreasing with decrease of the overall size. (Last property reveals  itself  in the $x$  dependences of parton distributions, in the special high energy processes, etc.)

Difference in the fluctuations of the wave function of a bound and free nucleon leads 
to the deformation of its wave function.  This phenomenon is well known  in  atomic physics 
from the application of the variational principle to the hydrogen molecule, $H_2$.
At large inter  proton distances the main effect is   swelling of the  hydrogen WFs 
while at small interproton distances the overall size of the hydrogen atoms is reduced \cite{Slater}.

The  use of the variational principle allows us to understand the qualitative trend in QCD as well 
  -- probabilities of the quark-gluon configurations  within a bound nucleon  for which attraction is
weaker than in average, should be reduced, while probabilities of configurations for which attraction is 
  stronger than in average should be enhanced.
 (This  is a particular case of the application of the Le Chatelier's principle.)

In \cite{Frankfurt:1988nt} the expressions were derived for the reduction of the probability of  
the configurations  in bound nucleons which interact with strength much smaller than average:
\begin{equation}
\langle \delta\rangle = 1 +{4\langle U\rangle\over \Delta E},
\label{overall}
\end{equation}
where $\langle U\rangle$ is the expectation value of the potential acting on a nucleon in the nucleus, $\Delta E = M_{N^*} - M_N\sim 400 \div 600 MeV$ is characteristic energy of excitation of the nucleon. Hence for a 
heavy nucleus a maximal suppression is of the order 20\%.  The calculation of the suppression effect for the case of 
quark-gluon
configurations for which strength of interaction is smaller but still comparable to  $\langle U\rangle$ requires use of a specific model, see
 \cite{Frank:1995pv}.

One can use equations of motion to derive the dependence of the suppression on the momentum of the bound nucleon \cite{Frankfurt:1985cv}. 
In the lowest order in $k^2/ m_N \Delta E$ neglecting term $\propto \epsilon_A/\Delta E$ one finds 
\begin{equation}
\delta(k)=1-2 k^2/m_N\Delta E.
\label{delta}
\end{equation}

Eq.\ref{delta} including the binding term can be written in a compact form if one substitutes $k^2$ in Eq.\ref{delta} by 
\begin{equation}
\Delta m^2= m^2_N - (p_A - p_{rec})^2.
\end{equation}
where $p_{rec}$ is the four momentum of the A-1 nucleon recoil system in the process where a nucleon was removed from the nucleus.  The possibility to rewrite Eq.\ref{overall}   in this form was first pointed out in \cite{Melnitchouk:1996vp}.
 A detailed analysis which demonstrated validity of the formula for the case of  generic final state of the nucleus was 
 performed in \cite{CFKS}. Note here that Eq.\ref{delta} naturally satisfies the requirement that 
 the scattering amplitude when continued to the pole where
 $\Delta m^2=0$, should coincide with the on-shell amplitude. Hence the linear dependence on off-shellness for small $\Delta m^2$ should be valid for wide range of bound nucleon deformation effects. In particular such a pattern is consistent with the measurements of the deviation of the ratio $G_E^{bound}/G_M^{bound}$  from the free value which was studied in the JLab experiment \cite{Malace:2010ft}.

\subsection{Implications for nuclear pdfs at large $x$}

The studies of the $x$-dependence of nucleon pdfs find a very different dependences on large $x$:     
$xV_N(x,Q^2)\propto (1-x)^3$, $x\bar{q}_N(x,Q^2)\propto (1-x)^7$ and $xG_N(x,Q^2)\propto(1-x)^5$.  This 
dependence indicates that 
hard interaction with nucleon at  $x\ge 0.6$  selects quark - gluon configurations in which antiquarks in the nucleon  and in particular the meson field  is suppressed.   The suppression of the interaction of nucleons in such configurations can be demonstrated   in the perturbative QCD  and on the level of emission of mesons as suppression of the pion emission by a small size configurations in a nucleon (see analysis in Appendix D of \cite{Frankfurt:1988nt}).

At the same time the analyses of low energy nucleon - nucleon interaction suggest   that
the dominant contribution to the  SRC  for $k\le 600$ MeV/c originates from the $I=0,S=1$ channel
where pion exchange plays a very important role (see e.g.  \cite{Kaiser:1998wa}). 

Combining this observation with the arguments of the previous subsection we conclude that probability of configurations with large $x$ partons should be suppressed in the bound  nucleons with the amount  of the suppression comparable to the one given by Eq.\ref{overall} which is of correct magnitude to explain the EMC effect for $x\sim 0.6\div 0.7$.

In the limit when one can average over the Fermi motion of nucleons ($x\le 0.6$) we find for the A-dependence of 
$\left <\delta \right> $:
\begin{equation}
1- \langle\delta \rangle = \langle 2k^2/\Delta E\rangle \propto T_A/m_N \propto a_2(A)
\end{equation}
Hence we conclude that for the discussed kinematics the hadronic EMC effect should 
be proportional to $a_2(A)$ with a good accuracy  which is indeed the case, see the discussion in section 6.

\subsection{Zooming on the nucleon polarization}

The critical test of the discussed picture would be a study of the
dependence of the nucleon deformation on the nucleon momentum which 
for a large range of nucleon momenta should be proportional to the square of the momentum of the struck nucleon.

The key experiment to test the discussed interpretation would be a measurement of the tagged EMC effect \cite{Frankfurt:1985cv},  the process 
$e +^2H \to e+ N + X$   
 with detection of the recoil nucleon with light cone fraction $\alpha > 1$. In this process one can  measure the bound 
 nucleon structure function $F_{2N}^{bound}(x/\alpha)$. We expect that 
\begin{equation}
1- F_{2N}^{bound}(x/\alpha,Q^2)/F_{2N}(x/\alpha,Q^2)= f(x/\alpha,Q^2) \delta m^2 ,
\end{equation}
with $f(x/\alpha,Q^2)$  
small 
for $x/\alpha < 0.5$ and rapidly growing  at larger $x/\alpha$. A fast 
dependence of the effect  
on $x/\alpha$ would make it easier to separate it from possible final state interaction effects
\footnote{In the case of the experiments planned at Jlab one would have to address the issue of the role of the higher twist effects for $F_{2N}(x\sim 0.5 ,Q^2).$}.

A priori the deformation of a bound nucleon
can also depend on the  angle $\phi$ between the momentum of the struck nucleon and the reaction axis as 
\begin{equation}
{{d\sigma\over d\Omega}\over \langle {d\sigma\over d\Omega}\rangle }=1+c(k,q).
\end{equation}
Here $\langle\sigma\rangle$ is cross section averaged over $\phi$ and 
$d\Omega$  is  the phase volume and the factor  $c$ 
characterizes non-spherical deformation. Correlation between the photon polarization and $\vec k$ direction is also possible.

Such non-spherical polarization  is well known in atomic physics.
 Contrary to  QED detailed calculations of this effect  are not possible in QCD.    However, a qualitatively similar deformation of the bound nucleons should arise  in QCD. One may expect that the  deformation of bound nucleon should be maximal in the  direction of radius vector between two nucleons of SRC.  \footnote{We are indebted to H.~Bethe(1995) and V.Gribov(1993) who draw attention of one of us (MS) to 
 this effect.}  The $\phi$-dependent  shape deformations of bound nucleons  are averaged out in the EMC effect  for $x < 0.7$.

As we mentioned above a non-spherical nucleon deformation may be manifested in the polarization effect in the process
$\vec e +^4He \to e+ \vec p +^3H$ where the the measured asymmetry can be interpreted in terms of the deviation of ratio $G_E/G_M$  for a bound nucleon from the free value.  It would be interesting to check whether the same pattern is observed for the scattering off the deuteron, whether the linear dependence on $\Delta m^2$ extends up to larger nucleon momenta and look for the dependence of the effect on the angle $\phi$. Since the deformation is expected to be strongest  along the radius vector between the nucleons, one may expect the deviations from the free nucleon case to be maximal for $\phi \sim 0,\pi$.

\section{Conclusions}
Application of  the baryon charge and momentum QCD sum rules allows us  to prove that the EMC effect is a genuine 
nuclear effect. To extract hadronic EMC effect from data one should take into account the  model independent effects. 
Additivity of the nucleus structure functions follows from nuclear physics ideas  if the  
Bjorken $x$   is used for the scattering off nuclei as dictated by the parton model and  QCD.   Account of QED requires to  include  the  equivalent photon component of the light-cone WF of the heavy nucleus. Comparison of nuclear and deuteron structure functions at the same variable $x_p$  leads to underestimate of the EMC effect for for all nuclei.
 At the same time the two effects we discussed mostly compensate each 
 other in the case of hadronic component of the EMC effect for  heavy nuclei.

  The data indicate that the hadronic EMC effect is a predominantly a high $x$ phenomenon. The A-dependence of 
 the hadronic EMC effect  is consistent with the A-dependence of the two nucleon correlations. The implementation  of 
 above mentioned effects impacts also on the nucleon pdfs extracted from the data and in particular 
 somewhat increases the nuclear correction for extraction of $F_{2n}(x,Q^2)$ from the deuteron data.

We also emphasized that the presence of predominantly nucleonic SRC in nuclei together with the similar A-dependence 
of SRCs and  the hadronic EMC effect,  and lack of the nuclear enhancement of antiquark distribution in nuclei,  put strong constraints on the mechanism of the EMC effect. These data  pose a serious problem for the mean field models where SRC 
are absent. At the same time they are consistent with the scenario of the nucleon deformation  occurring  predominantly due 
to suppression of small size,  large $x $ configurations in the bound nucleons. It is based on  the well established properties of bound states in QCD: presence of  fluctuations with  strength of the interaction 
and of quark-gluon content, and selection of the   certain instantaneous  quark-gluon configurations by hard processes.
Deformation of peripheral pion cloud of a nucleon is suppressed since in low energy QCD pion-pion interaction is suppressed by high power of pion momentum.

It is important to perform further experimental measurements of the EMC ratio for $x\ge 0.5$ at large $Q^2$ which will become feasible at JLab 12. Especially interesting would be to study the ratio of the 
EMC effect in $^{48}$Ca and $^{40}$Ca since in this case the standard model leads to the same nuclear effect.
  Moreover such research would  provide a unique window on the structure of neutron rich high density nuclear matter relevant for description of the cores of the neutron stars.

The critical tests of the current ideas will be possible in the processes where momentum of the struck nucleon is tagged. 
It is expected that effect should be proportional to the nucleon off-shellness and may also depend on the direction between the momentum of the struck nucleon and the virtual photon momentum (polarization).

It would also be  interesting to analyze the role of the violation of the  isotopic invariance due to presence of the  Coulomb field in nuclei and in particular the difference of the momentum fraction carried by photons in protons and neutrons ($\eta_{\gamma}(p)- \eta_{\gamma}(n) \approx 0.2 \%$)  and its role in the   extraction the Weinberg angle from neutrino data\cite{nutev}.

\section*{Acknowledgements}
We thank  W.Vogelsang for informing us  about   early studies of the effect of the Coulomb field in nucleons. 
We are indebted to V.Gribov(1993) and H.Bethe (1995)  who draw our attention  to angular asymmetry of 
deformation of atoms in QED in the close interaction.  We also thank C.~Ciofi degli Atti and L.~Kaptari for the 
collaboration studies of the color fluctuation model. We thank  V. ~ Guzey for pointing out the sign error in the original 
reconstruction of  $R_A(x,Q^2)$ from experimental data presented as a function of $x_p$  and help with generating several plots. We thank Or Hen  for discussion of the x-dependence of $F_{2N}$ in the kinematics of the JLab and SLAC experiments.  The research was supported by DOE and BSF.

\end{document}